\documentclass[conference,12pt,onecolumn]{IEEEtran}

\usepackage{cite}

   \usepackage[pdftex]{graphicx}

\usepackage[cmex10]{amsmath}
\usepackage{amssymb} 			

  \usepackage[caption=false,font=footnotesize]{subfig}
\usepackage{fixltx2e}
\usepackage{url}

\newcommand{\be}{\begin{equation}}
\newcommand{\ee}{\end{equation}}

\newcommand{\qed}{\nobreak \ifvmode \relax \else
      \ifdim\lastskip<1.5em \hskip-\lastskip
      \hskip1.5em plus0em minus0.5em \fi \nobreak
      \vrule height0.75em width0.5em depth0.25em\fi}
      
\usepackage{listings}
\usepackage{color}

\begin{document}
%
\title{Model Predictive Frequency Control \\Employing Stability Constraints}
%
%
%

\author{
	Christoph Trabert, Andreas Ulbig and G\"{o}ran Andersson\\
	Power Systems Laboratory, ETH Zurich\\
	Zurich, Switzerland\\
	{\textmd ctrabert@ee.ethz.ch$\;$\textbar$\;$}{\textmd \{ulbig$,\;$andersson\}@$\,$eeh.ee.ethz.ch}
}
%
%


\markboth{Journal of Power System Stability for Dummies,~Vol.~1, No.~1, January~2014}
{Trabert \MakeLowercase{\textit{et al.}}: Model Predictive Frequency Control Employing Stability Constraints}

%


\IEEEspecialpapernotice{(Technical Report)}

\maketitle

\begin{abstract}
	In this article a model predictive control (MPC) based frequency control scheme for energy storage units was derived, focusing on the incorporation of stability constraints based on Lyapunov theory and the concept of passivity.

The proposed control schemes, guaranteeing closed-loop stability, are applied on a one-area and two-area power system. For the two-area power system, a coordinated (centralized) control and an uncoordinated (decentralized) control approach is conducted.

The stability properties of the different MPC setups were derived, implemented and simulated. Furthermore the corresponding control performance was analyzed.

\end{abstract}


%
\IEEEpeerreviewmaketitle

\section{Introduction}

\subsection{Motivation}
\IEEEPARstart{T}{raditionally}, power system operation has essentially been based on the assumption, that electricity is reliably and steadily produced by large power plants, which are fully dispatchable, i.e. controllable and have a high frequency inertia.\\
However, a strong trend towards generation of electrical energy by renewable energy sources (RES), i.e. PV units and wind turbines with no or decoupled rotating masses, exists\footnote{On 16th of June 2013, the share of energy produced by wind and solar power reached a level of above 60 \% for the first time in Germany. It resulted in negative electricity prices in France and Germany \cite{res_infeed_germany}.}. How power systems could be adapted in order to accommodate for increasing shares of uncontrolled fluctuating RES as well as power market activities is a highly relevant and still open research question.\\
Options to deal with these challenges might be the expansion of transmission capacities, the extensive integration of storage capacities as well as a better exploitation of the inherent flexibility within the power system.\\
Model predictive control (MPC) as an optimal control scheme for regulating grid frequency receives rising attention due to rapidly growing shares of variable RES and thereby arising challenges for power system operation. \\
The choice of MPC as a control approach is especially motivated by its ability to incorporate operational constraints of power systems, which cannot be handled by conventional P/PI-controllers. It enables the provision of frequency control using a generic power system storage unit, e.g. a battery, with given operational constraints, such as the power ramp rate, power rating and storage capacity.
Additionally, the recently growing interest in using MPC for control purposes of power systems emerges due to the availability of faster and cheaper computing resources, as a significant computational effort comes with the use of such a receding horizon control scheme.

\subsection{Goal and Content}

The objective of this article is to derive an MPC based frequency control scheme for energy storage units. Specifically, the employment of stability constraints will be studied. The theoretical focus lies on the derivation of passivity as a concept for guaranteeing stability. Furthermore stability guaranties derived from Lyapunov-Theory, so called Control Lyapunov Functions (CLF) are incorporated into the optimal control setup.\\
Eventually, the goal is to ensure a stable frequency regulation of a one-area and a two-area power system, which is disturbed by a fault signal. 

The remainder of the article is organized as follows: Section \ref{sec:modelling} elaborates on the dynamics of grid frequency. Subsequently in section \ref{sec:conventional}, first, a theoretical introduction to stability is stated, followed by derivation, simulation and validation of conventional control. 
Subsequently in section \ref{sec:mpc_mpfc}, the derivation of the model predictive frequency control (MPFC) problem is given. In section \ref{sec:guaranteedstability} different approaches to ensure stability of the controller are derived. This is followed in section \ref{sec:study_case_results} by a study case. Finally, the outcome of the article is presented in section \ref{sec:conclusion_outlook}.  \\

%
%
%
%

%
%
%

\section{Power System Modeling and Analysis}
\label{sec:modelling}


\subsection{Grid Frequency}
\label{sec:grid_frequency}

Power systems are dynamic systems with a high degree of complexity. Corresponding processes within the power system therefore happen on several time-scales (milliseconds to yearly seasons), with a large spatial distribution (hundreds to thousands of kilometers) and numerous grid hierarchy levels (voltage levels). \\
Dominant for the effect of frequency stability, and hence stable operation, is active power balance, meaning that the total power in-feed minus the total consumption is zero. The nominal frequency is set to $f_0 = 50$ Hz as in the ENTSO-E Continental Europe system (formerly UCTE).
Generally deviations from the nominal frequency arise from imbalances between instantaneous electric power consumption and production. If there is a higher consumption than production in a time instant, this results in a decelerating effect on the synchronous machines. On the contrary, higher production than consumption of electric power, leads to an accelerating effect on the synchronous machines \cite{psd_andersson}.\\ 
Small local disturbances can evolve into consequences influencing the whole power system, which could lead to cascading faults and black-outs in the worst case. \\
The key system state to observe is the grid frequency $f$, concerning this (dominant) aspect of power grid stability. As the rotational speed of the synchronous generators  $\omega$ is inherently coupled to grid frequency ($\omega=2\pi f$), frequency deviations should be kept as small as possible, since this leads to damaging vibrations in synchronous machines. \\
Maintaining the grid frequency within an acceptable range is therefore required for stable operation of the power system. Small variations occur spontaneously by small load or generation deviations and usually do not have critical consequences in normal operation.\\
Large frequency variations, which might be caused by errors in demand forecasts or RES forecasts, the spontaneous loss of load or generation units, however, could lead to situations, where synchronization between the generating unit is lost. This happens when the angle differences get too large, which would result in (possibly cascading) disconnection of the machines \cite{paper_rinke}.

\subsection{Aggregated Swing Equation}
\label{sec:ASE}

The frequency dynamics of a power system can be described by the aggregated swing equation (ASE) 

\begin{equation}\Delta\dot{f}=\frac{f_0}{2H}\left(\Delta P_{m}(p.u.) - \Delta P_{\textrm{load}}(p.u.)\right), \label{eq:ase} \end{equation}

with $\Delta P_{\textrm{load}}$ being the deviation of load, $\Delta P_m$ being the mechanical (or inert) power deviation, and $H$ being the total inertia constant\footnote{$H$ is typically valued in the range 2...10 s, cf. \cite{kundur1176power}.}.

Please note, that losses are not taken into account here. Regarding the stability analysis, which is done in the sequel, this is a very conservative assumption.

The following assumptions are made

\begin{itemize}
	\item The synchronous machine is modeled as a constant electro-magnetic field behind the transient reactance. The angle of the electro-magnetic field is assumed to coincide with the rotor angle.
	\item Resistances in lines, transformers and synchronous machines are neglected.
	\item Voltages and currents are assumed to be perfectly symmetrical, i.e. pure positive sequence.
	\item The angular velocity is close to nominal.
	\item Static models for lines are used.
\end{itemize}

Despite its simplicity, a number of important conclusions concerning the angular stability in large systems can be drawn from this analysis \cite{psa_andersson}.

\subsection{Frequency Dependency of the Loads}
\label{sec:freq_dep_load}
In real power systems, a frequency dependency of the aggregated system load is clearly observable. This has a stabilizing ("self-regulating") effect on the system frequency $f$.

It is assumed, that the damping power of the system can be written as
\begin{equation} P_d = \frac{1}{D_{l}}\cdot \Delta f  \nonumber. \end{equation}


Including the frequency-dependency of the loads into equation \ref{eq:ase} leads to

%

\begin{equation}\Delta\dot{f}=-\frac{f_0}{2HD_{l}S_B} \Delta f + \frac{f_0}{2HS_B}\left(\Delta P_{m} - \Delta P_{\textrm{load}}\right). \label{eq:ase_fd}\end{equation} 

Rotating mass loads are neglected. They play a decreasing role in future power systems and are not of big importance for (conservative) stability analyses. \\

\subsection{Mathematical Modeling}
\label{sec:mathematical_modeling}

For the system dynamics the following basic model is assumed:
\begin{equation} \dot{x}=f\left(x,u\right), \label{eq:nonlinear}\end{equation}
where $x = \begin{bmatrix} \Delta f & x_{\textrm{SoC}} \end{bmatrix}^T,$ with $\Delta f \in \mathbb{R}$ referring to the deviation of the normal frequency of the grid and $x_{\textrm{SoC}} \in [-1,1]$ to the state of charge of a connected battery.\\

The corresponding linearization in $\left(x,u\right)=\left(x_{0},u_{0}\right)$ as a state space model is given by: 
\begin{eqnarray}
\dot{x} & = & \left.\frac{\partial f}{\partial x}\right |_{x_{0},u_{0}} x+\left.\frac{\partial f}{\partial u}\right|_{x_{0},u_{0}}u\label{eq:linsys}\\
 & = & A_{c}\left(x_{0},u_{0}\right)x+B_{c}\left(x_{0},u_{0}\right)u\nonumber \\
  & = & A_{c}x+B_{c}u. 
\end{eqnarray}

%
Next, the continuous linear system is discretized
\begin{equation} x_{k+1} = A_{d}x_{k}+B_{d}u_{k}\label{eq:disclinsys}.\end{equation}

To be able to account for energy constraints of the control input later (due to energy storage restrictions etc.), the additional power $u = P_{\textrm{add}}$ is integrated as following, and  $x_{\textrm{SoC}}$ is used as another state within the state space model:

\begin{equation} E_{\textrm{add}} = \int P_{\textrm{add}} \;\mathrm{d}t = \int -C_{\textrm{bat}}\dot{x}_{\textrm{SoC}} \;\mathrm{d}t.\nonumber \end{equation}

Including the possibility of losses, respectively self-discharging $v$ of the battery with capacity $C_{\textrm{bat}}$ leads to 
\begin{equation} \dot{x}_{\textrm{SoC}} = -\frac{v}{C_{\textrm{bat}}}-\frac{1}{C_{\textrm{bat}}}\cdot u, \label{eq:soc}\end{equation} with $x_{\textrm{SoC}} \in [-C_{\textrm{bat}}/2,C_{\textrm{bat}}/2 ]$. Combining equation \ref{eq:soc} and the aggregated swing equation \ref{eq:ase_fd} and transforming it into (continuous) state space form yields:

\begin{equation} \frac{d}{dt} \begin{bmatrix} \Delta f \\x_{\textrm{SoC}} \end{bmatrix} = \begin{bmatrix} A_{\textrm{freq}} & 0 \\ 0 & \frac{-v}{C_{\textrm{bat}}}\end{bmatrix}\begin{bmatrix} \Delta f \\x_{\textrm{SoC}} \end{bmatrix} + \begin{bmatrix}B_{\textrm{freq}} & 0 \\ 0 & \frac{-1}{C_{\textrm{bat}}} \end{bmatrix} \begin{bmatrix} \Delta P \\ u \label{eq:linsys} \end{bmatrix} \end{equation}

with:
\begin{align*}
A_{\textrm{freq}}  =  \frac{-f_0}{2HS_{B}D_{l}},\qquad
B_{\textrm{freq}}  =  \frac{f_0}{2HS_{B}}.
\end{align*}

$\Delta P$ refers to the deviation of the expected power, including the fault power, therefore, it directly influences the frequency deviation in the grid.\\
$u$ however refers to the electrical power, which is injected into the grid by the controller, hence directly influencing the state of charge of the battery \cite{general_fc_ulbig}, \cite{powernodes_ulbig}.

\subsection{Two-Machine Model}
\label{sec:two_machine_model}
\begin{figure}
	\centering
	\includegraphics[trim = 0cm 0cm 0cm 0mm, clip, width=0.6\textwidth]{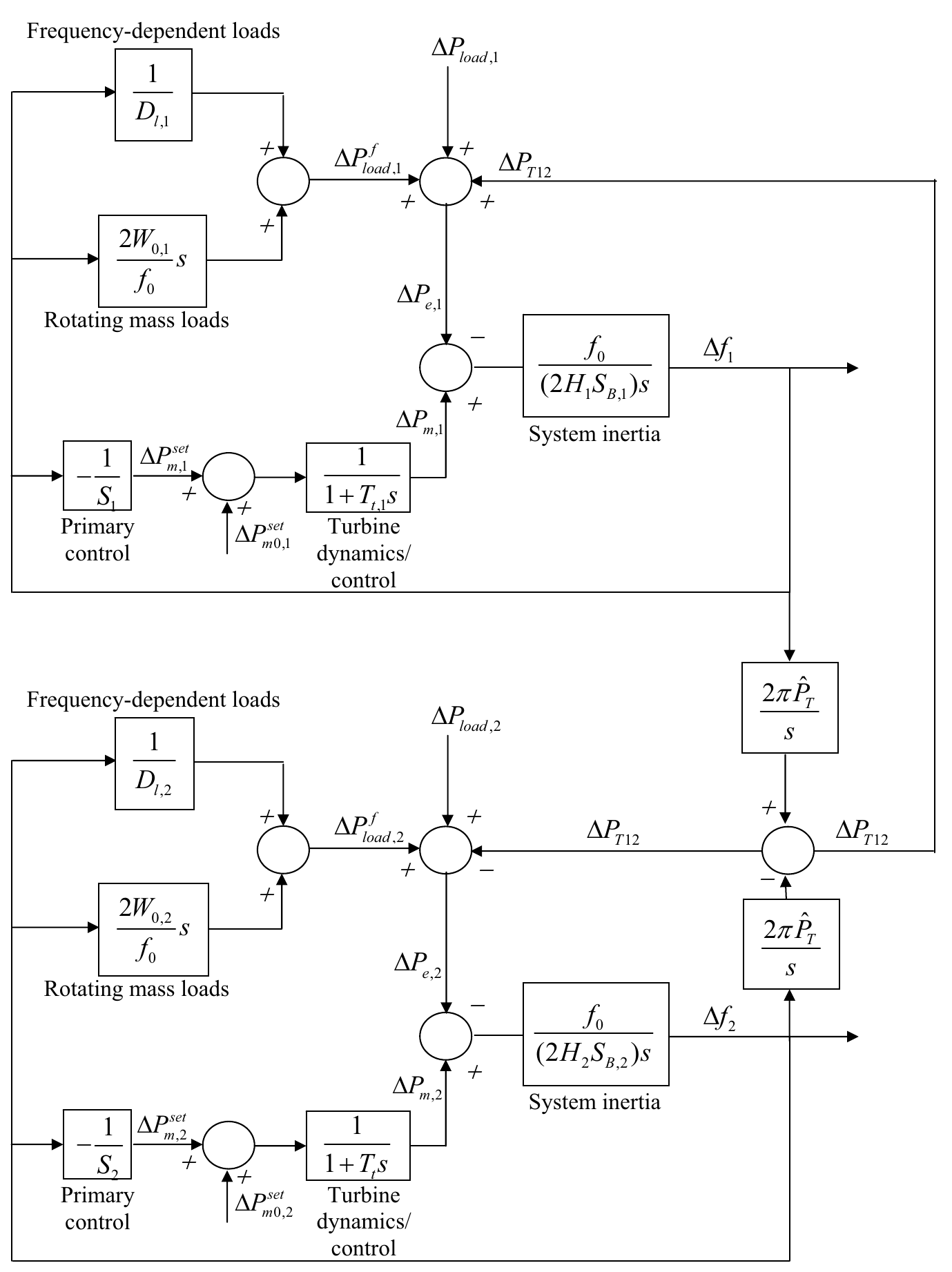}
	\caption{Two-machine respective two-area model including tie flows (from \cite{psd_andersson}).}
	\label{pic:TwoAreaSystem}
\end{figure}

If a power system is highly meshed, it can be seen as all units being connected to one single bus, i.e. acting as one (aggregated) swing equation. However, in practice, a large interconnected system is divided into several "control zones". To understand the interaction between two such zones, in this section a two-area model as in figure \ref{pic:TwoAreaSystem}, with a tie-line in between is derived.

The power exchange between the two zones is
\begin{equation} P_{T12} = \frac{U_1 U_2}{X}\sin(\varphi_1 - \varphi_2), \nonumber \end{equation}
where $X$ equals the (equivalent) reactance of the tie line between zone 1 and zone 2. This can also be written as:
\begin{equation} P_{T12} = \hat{P}_T\sin(\varphi_1 - \varphi_2), \nonumber \end{equation}

with $\hat{P}_T$ being the maximum power transmission
\begin{equation} \hat{P}_T = \frac{U_1 U_2}{X}. \nonumber \end{equation}

The (aggregated) swing equation divides into a system of equations, comprising the interaction between the two areas:
\begin{eqnarray}
\Delta\dot{f}_1 & = & \frac{f_0}{2H_1 S_{B,1}}\left(\frac{-1}{D_{l,1}}\Delta f_1 - \hat{P}_T\sin(\varphi_1 - \varphi_2) + u_1\right),\nonumber \\
\Delta\dot{f}_2 & = & \frac{f_0}{2H_2 S_{B,2}}\left(\frac{-1}{D_{l,2}}\Delta f_2 + \hat{P}_T\sin(\varphi_1 - \varphi_2) + u_2\right), \nonumber \\
\Delta \dot{\varphi} & = & \frac{d}{dt} (\varphi_1 - \varphi_2) = 2 \pi (\Delta f_1 - \Delta f_2).
\label{eq:2ase} 
\end{eqnarray}

After linearizing the model around the set point $ x = \begin{bmatrix}
0 & 0 & 0 & 0 & 0 \end{bmatrix}^T $, where $\sin(\varphi_1 - \varphi_2) \approx (\varphi_1 - \varphi_2) = \Delta \varphi$, the state space matrices change to

\begin{equation} A_{\textrm{coupled}} = \begin{bmatrix} 
A_{\textrm{freq},1} & 0 & 0 & 0 & A_{\textrm{freq},1} D_{l,1}\hat{P}_T \\
0 & \frac{-v_1}{C_{\textrm{bat,1}}} & 0 & 0 & 0\\ 
0 & 0 & A_{\textrm{freq},2} & 0 & - A_{\textrm{freq},2}D_{l,2}\hat{P}_T \\
0 & 0 & 0 & \frac{-v_2}{C_{\textrm{bat,2}}} & 0\\  
2\pi& 0 & -2\pi & 0 & 0\end{bmatrix},\nonumber \end{equation}
and
\begin{equation} B_{\textrm{coupled}} = \begin{bmatrix} 
B_{\textrm{freq},1} & 0 & 0 & 0 \\ 
0 & \frac{-1}{C_{\textrm{bat,1}}} & 0 & 0 \\
0 & 0 & B_{\textrm{freq},2} & 0 \\
0 & 0 & 0 & \frac{-1}{C_{\textrm{bat,1}}} \\ 
0 & 0 & 0 & 0 \\
\end{bmatrix}, \nonumber\end{equation}
with
\begin{align*} 
A_{\textrm{freq,}i}  =  \frac{-f_0}{2H_i S_{B,i}D_{l,i}}, \qquad
B_{\textrm{freq,}i}  =  \frac{f_0}{2H_i S_{B,i}},
\end{align*}

%

the corresponding states $ x = \begin{bmatrix}
\Delta f_1 & x_{\textrm{SoC},1} & \Delta f_2 & x_{\textrm{SoC},2} & \Delta \varphi
\end{bmatrix}^T $ and inputs $ u = \begin{bmatrix}
 \Delta P_1 & u_1 & \Delta P_2 & u_2 \end{bmatrix}^T. $

\subsection{Simulation of Power System Dynamics}

For simulation purposes, power system parameters as in table \ref{tab:parameters} were assumed. For load damping, the minimum measurement in \cite{weissbach2008improvement} was assumed, to carry out very conservative stability analysis. 

\begin{table}[h]
\centering
\renewcommand{\arraystretch}{1.5}
\begin{tabular}{l|l|l}
Unit & Size & Description \\
\hline 
\hline
$H$ & 6 s & system inertia \\ 
\hline 
$k_{\textrm{pf}}$ & $1.5$ $\%$ & self regulating effect \\ 
\hline 
$D_l$ & $\frac{1}{k_{\textrm{pf}}} = 66.67$ $\%^{-1}$ & load damping \\ 
\hline 
$S$ & $\frac{200\textnormal{ mHz}}{3000\textnormal{ MW}} $ & frequency droop, for primary freq. control\\ 
\hline 
$T_N$ & 240 s & integration param. for conv. freq. control \\ 
\hline 
$C_P$ & 0.17 & proportional param. for conv. freq. control \\ 
\hline 
$B$ & $\frac{1}{D_l}+\frac{1}{S}$ = 20550 & bias factor, for conv. control \\ 
\hline 
\end{tabular} 
\vspace{0.3cm}
\caption{Power system parameters used.}
\label{tab:parameters}
\end{table}

\subsubsection{Fault Signal}
\label{sec:fault_signal}
\begin{figure}
	\centering
	\includegraphics[trim = 2cm 7.1cm 1cm 8.5cm, clip, width=0.45\textwidth]{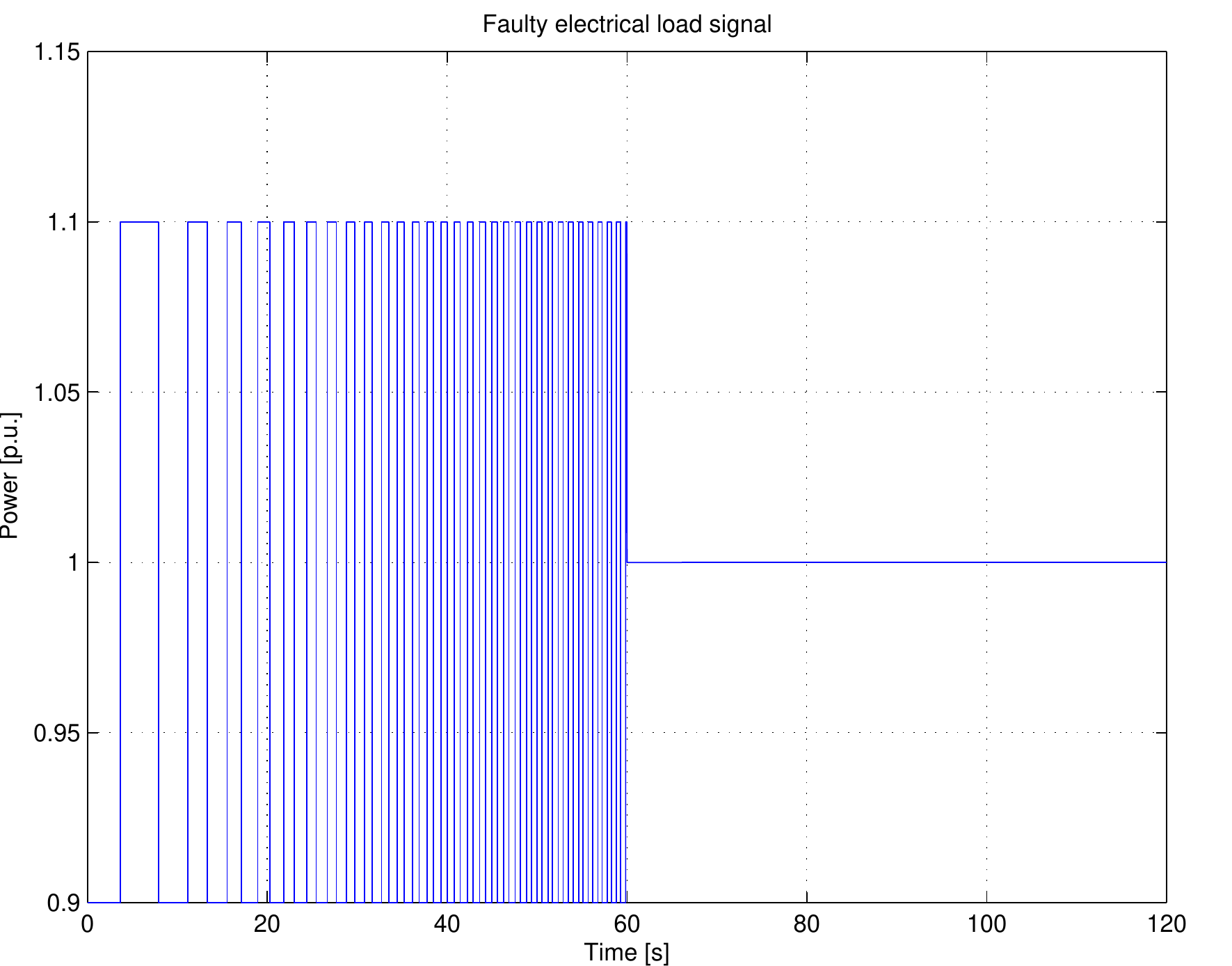}
	\caption{Used fault signal, note asymmetric behavior: upper half-waves are shorter than the lower ones (resulting in a decreasing moving average).}
	\label{pic:Faulty_Chirp}
\end{figure}

\begin{figure}
	\centering
	\includegraphics[trim = 2cm 7.1cm 1cm 8.7cm, clip, width=0.45\textwidth]{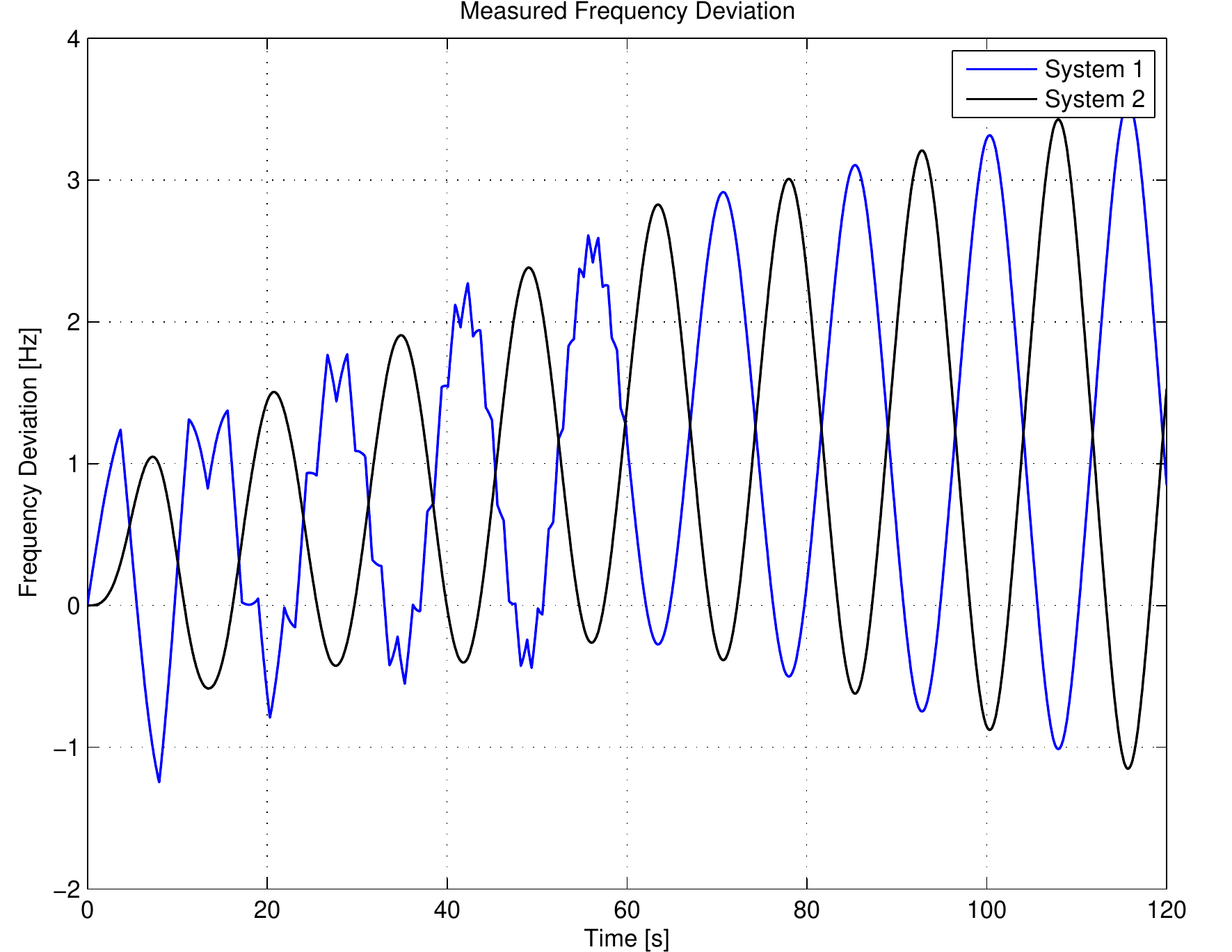}
	\caption{Uncontrolled frequency response on given fault signal on a two area system.}
	\label{pic:freq_dev_uncontrolled_chirpfault_2sys}
\end{figure}

In all simulations, as long as not stated otherwise, the disturbance signal given in the power deviation graph in figure \ref{pic:Faulty_Chirp} is introduced in the power system. This signal was taken as it provokes instabilities for (conservative) stability tests due to enormous sudden disturbances, e.g. cascading disconnection of load or generation units. 
Applying the fault to an uncontrolled, system comprising 
two control areas shows the frequency response in figure \ref{pic:freq_dev_uncontrolled_chirpfault_2sys}.

%

%
%
%

\section{Conventional Frequency Control}
\label{sec:conventional}


\subsection{Power System Stability}
A dynamic phenomenon in a power system is initiated by a disturbance in the system, e.g. a a line or a generator trips. While smaller disturbances usually result in small transients in the system, which diminish quickly because of damping, larger disturbances could have a significant impact and excite larger oscillations.
Hence, stability is associated with decaying oscillations and that the operation of the power system can continue without major impact for any power consumer \cite{psa_andersson}.


In this article the focus lies on frequency stability, corresponding to the frequency dynamics, which were analyzed in section \ref{sec:modelling}. Unlike voltage stability, which only acts locally, frequency stability plays a global role in a power system. It refers to the difference between the total power fed into the power system and the total power consumed by the loads, including the losses. If the resulting imbalance is comparatively small, the generators participating in frequency control will regulate the active power input from their prime movers, and bring back the frequency deviation to acceptable values. But if the imbalance is too large, the frequency deviation gets significantly big and might cause serious consequences.
In real power systems, instabilities result both from active and reactive power imbalances, so the assumptions, which are made in this article, are of course not valid in every case of power system instabilities. However, in many cases it is possible to identify active power imbalances and resulting frequency instability as the dominating processes at the start of power system instabilities \cite{psa_andersson}.

\subsection{Conventional Frequency Control}

The nominal grid frequency is defined to be $f_0=50$ Hz in the grid zones of the European Network of Transmission System Operators for Electricity (ENTSO-E). To maintain this frequency, for secure operation, electricity demand and supply have to be in balance at every point in time. Small local disturbances in a power system can lead to consequences influencing the whole system, in the worst case ending in a black-out.

\begin{figure}
	\centering
	\includegraphics[trim = 0cm 0cm 0cm 0mm, clip, width=0.45\textwidth]{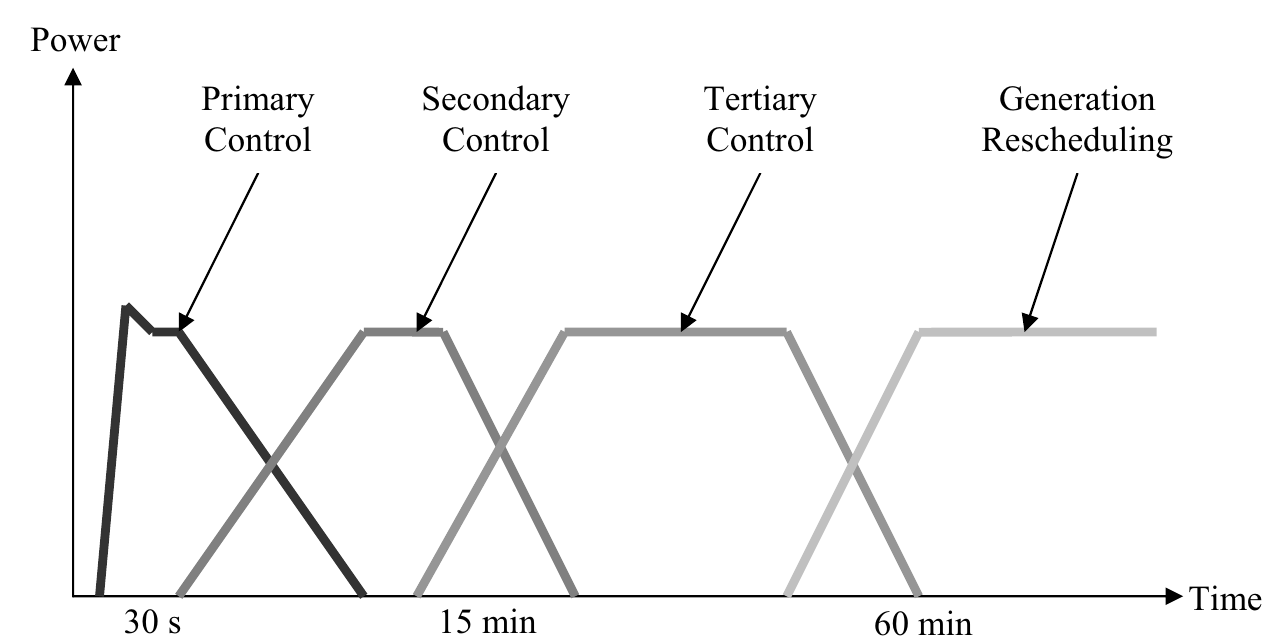}
	\caption{Different categories of frequency control over time (from \cite{psd_andersson}). Primary and secondary control are referred to as load frequency control (LFC), while tertiary control and generation rescheduling are called economic dispatch (ED).}
	\label{pic:ConventionalControl}
\end{figure}

In order to secure a stable operation, a frequency regulation mechanism is implemented. In traditional frequency  control, there are three categories (cf. figure \ref{pic:ConventionalControl}): \emph{Primary frequency control} provides power output proportional to the deviation of the system frequency, which stabilizes the system without restoring the reference frequency. Its time horizon lies within a few seconds after the occurrence of the deviation. Furthermore \emph{secondary frequency control}, taking over after approx. 30 s, has an integral control part, which restores the reference frequency. The time constant of secondary frequency control is chosen to be significantly larger than the time constant for primary control in order not to interfere with it as well as avoid "wear and tear" for the units.  \emph{Tertiary control}, finally, is operated manually and adjusts power generation set-points about 15 min after severe faults. \emph{Generator rescheduling} is dispatched through intra-day auctions, according to the estimated permanent fault magnitude, aiming to relieve tertiary control by cheaper sources \cite{paper_rinke}.



One of the factors used in this control scheme is the \emph{frequency bias factor}. Load frequency control is based on the non-interaction principle, hence, the disturbance power balance between all neighboring areas is restored \cite{biasfactor}. 

Referring to the system self-regulating effect (cf. section \ref{sec:freq_dep_load}) in the European power system, a frequency dependency of at least
$ k_{\textrm{pf}} = \frac{1}{D_l^{p.u.}} =  \frac{\Delta P / P_0}{\Delta f / f_0} = 1.5 \frac{\%}{\%}$
was estimated in \cite{ifac_welfonder}.

\subsection{Simulation of Conventional Frequency Control}

\paragraph{One Area}

\begin{figure}
\begin{minipage}[hbt]{0.45\textwidth}
	\centering
	\includegraphics[trim = 2cm 7.1cm 1cm 8.7cm, clip, width=1\textwidth]{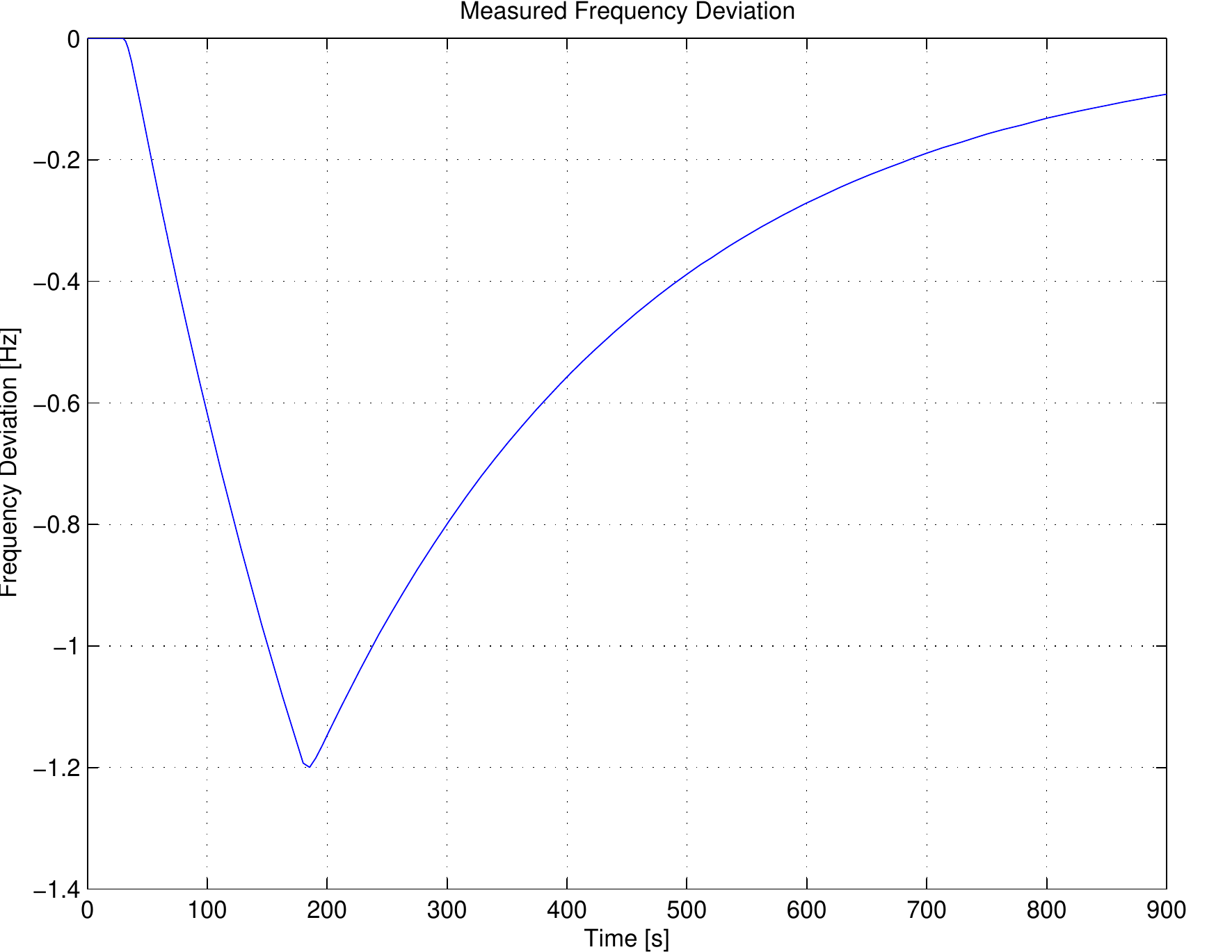}
	\caption{Frequency response of the developed power system model (one area) controlled by  conventional P/PI-control loops.}
	\label{pic:ConventionalControl_Rampe}
\end{minipage}
\hfill
\begin{minipage}[hbt]{0.45\textwidth}
	\centering
	\includegraphics[trim = 2cm 7.1cm 1cm 8.7cm, clip, width=1\textwidth]{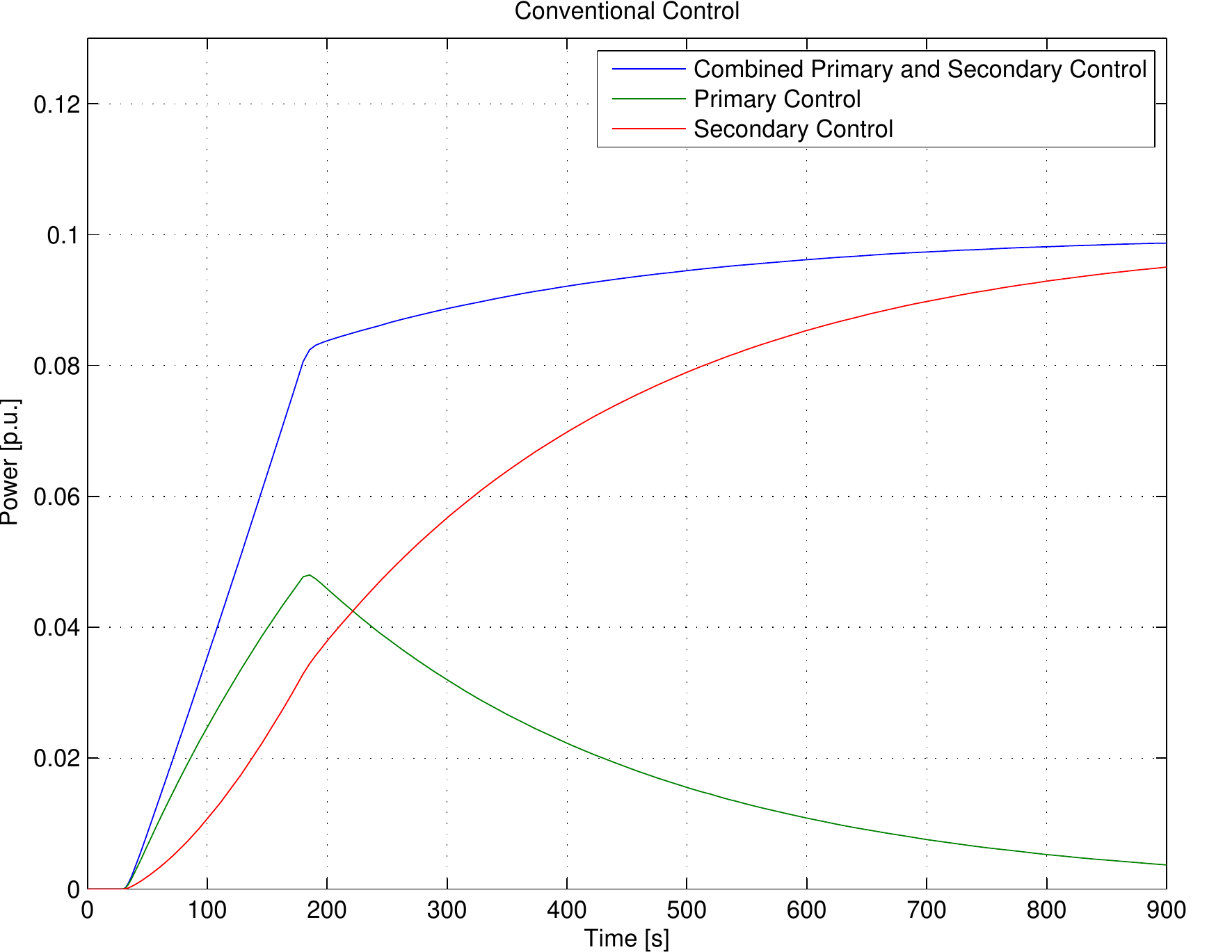}
	\caption{The conventional primary and secondary control input, which yields to the frequency deviation in figure \ref{pic:ConventionalControl_Rampe}.}
	\label{pic:PrimSecondCombiControl_conventional}
\end{minipage}
\end{figure}

The simulation of the frequency response in figure \ref{pic:ConventionalControl_Rampe} was conducted to validate the developed model. Together with the conventional primary and secondary (cf. figure \ref{pic:PrimSecondCombiControl_conventional}), it shows very similar behavior as in other publications and power grid measurements (\cite{weissbach2008improvement}).


\paragraph{Two Areas}
In this setup, two equally sized areas are interconnected via a tie-line, e.g. with a transmission capacity of $\hat{P}_T=0.2$~p.u. Each controlled area is equipped with an individual conventional primary and secondary controller. \\

\begin{figure}
	\centering
	\includegraphics[trim = 2cm 7.1cm 1cm 8.7cm, clip, width=0.45\textwidth]{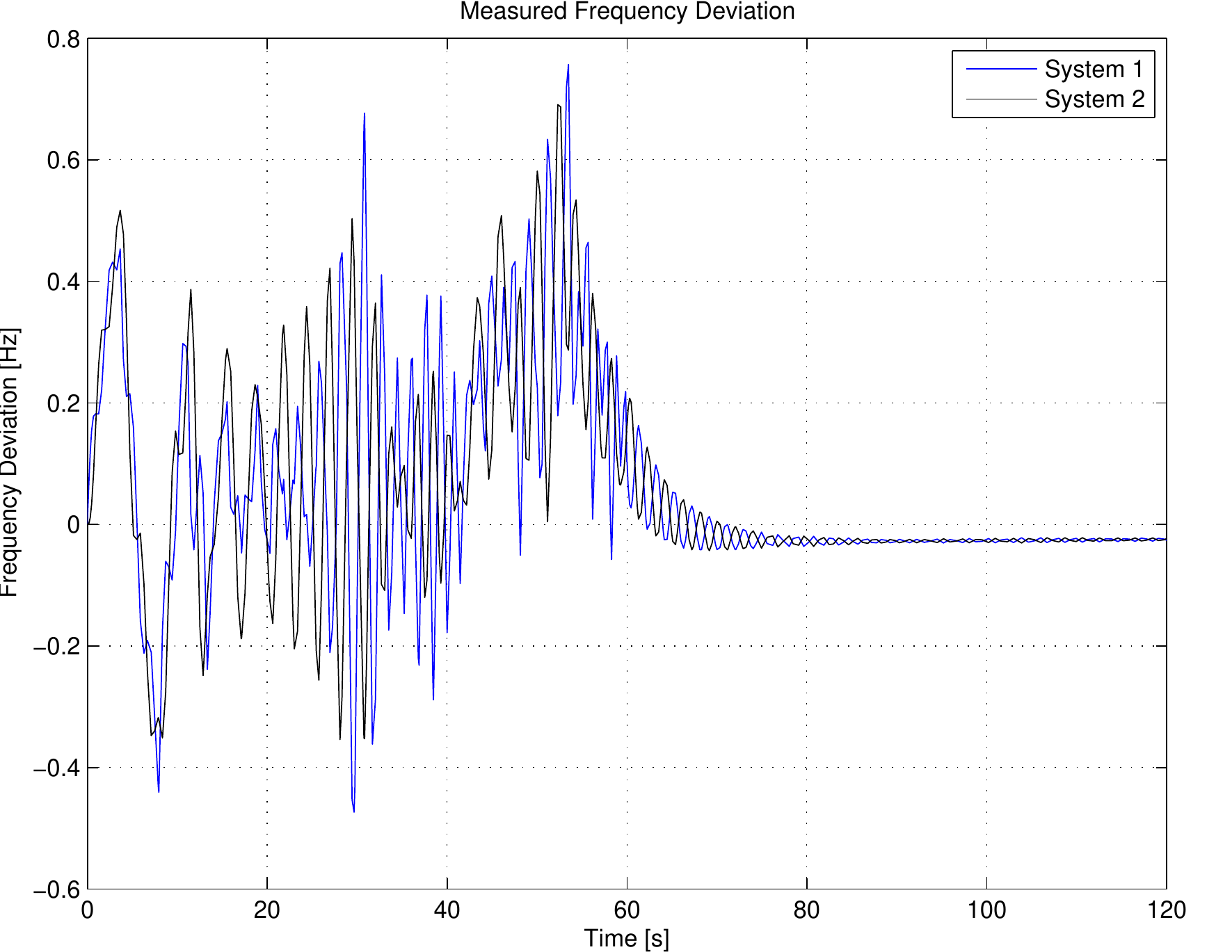}
	\caption{Frequency response of the developed model with two interconnected power systems controlled by two individual conventional P/PI-controllers. Fault signal as before.} 
	\label{pic:FreqDeviation_conventional_faultySignal1_2sys}
\end{figure}


\subsection{Motivation for new Methods}

In conventional systems, primary and secondary frequency control, referred to as automatic generation control (AGC), usually is implemented with satisfactory results. \\

The main conventional AGC properties are (summarized from \cite{carpentier1985or}):
\begin{itemize}
	\item \emph{"The control system is naturally area-wise decentralized."} Except in the case of a lack of area generation capacity, the frequency deviation diminish and all the inter-area power transmissions return to their preassigned values, while each area needs only data available within the area itself. 
	\item \emph{"The control system is robust."} Robustness is meant in a "rustic, simple and not fragile way": 
	\begin{itemize}
		\item Only regional data is needed.
		\item Only little information is needed: frequency, the sum of exported powers and the electric generated powers, with possibly low sampling rates and large transmission delays.
		\item It is practically insensitive to the exact physical system parameter values as well as control system failures.
\end{itemize}	 
	
	\item \emph{"The control system is sufficiently stable in practice,"} if it is carefully designed and tuned. Both static and dynamic aspects have been studied in the past.
	\item \emph{"Adaptive control is easy to implement."} This is due to the system's simplicity.
	\item \emph{"Control may not be sufficiently smooth."} No attention is paid to unit response rate limits in the controller. In practice, regulating margins are used, but rate constraints can not be guaranteed to be met.
	\item \emph{"The interface between load frequency control\footnote{I.e. primary and secondary control.} and economic dispatch\footnote{I.e. tertiary control and generation rescheduling.} is not satisfactory."} Due to different time scales, the orders of the different control levels might contradict each other, resulting in undesired oscillations as well as useless wear of the generation units.
\end{itemize}

Besides the final two properties, the given points essentially explain the success and long life of the conventional implementations. \\ 

But, many proposals exist for modern control systems, most of them applying optimal control theory or optimal power flow techniques. The relevant capabilities and desirable properties of these modern proposals are seen as high (summarized from \cite{carpentier1985or}):
\begin{itemize}
	\item It could take system constraints into account, e.g. transmission security and power rate limits.
	\item It might improve economy, by an accurate application of economic dispatch and the reduction of regulating margins.
	\item It still must have good transients, with large stability margins and smooth control, but especially addressing the LFC-ED interface problem.
	\item It must be kept robust, i.e. not fragile, by avoiding data complexity, and, probably most important, being area-wise decentralized.
\end{itemize}

With all potentials lying in modern control techniques involving optimal control and optimal power flow: one should never forget the engineer's perspective, particularly regarding the objectives, the constraints and, especially, the robustness of the control system \cite{carpentier1985or}.

%
%
%

\section{Model Predictive Control}
\label{sec:mpc_mpfc}


\subsection{Main Concept of Model Predictive Control}
\label{sec:mpc}

The design of a stabilizing feedback with a performance criterion being minimized, while satisfying constraints on the control input and the system states at the same time, is desired for many control problems. 
A solution approach is to repeatedly solve an open loop optimal control problem. The first sample of the resulting open-loop optimal trajectory, is then applied to the real system. Afterwards the whole process is repeated for the next time step.
Control methods like these are referred to as either model predictive control (MPC), moving horizon or receding horizon control (RHC) \cite{survey_mpc}. 


MPC is formulated as a repeated online solution of a finite horizon open-loop optimal control problem. When measuring the initial system state $x(0)=\left[x_1(0), ..., x_n(0)\right]^T=x_0$, a control input over the prediction horizon $T_p$ can be determined, by solving the following optimal control task:
\begin{eqnarray}
\min\limits_u & & J_{k} \nonumber\\
\mbox{s.t.} &  x_{k+1} &= f(x_k,u_k)\nonumber\\
			&  u_{k}&\in\mathcal{U}\nonumber\\
			&  x_{k}&\in\mathcal{X}\nonumber\\
			&  k &=1,...,N \label{eq:mpc_general}\
\end{eqnarray}
with the control input $u \in \mathbb{R}^g$ and the system state $x \in \mathbb{R}^n$. $k$ is an integer counting in time: $k \cdot T_s \leq T_p$, $J_k$ refers to the cost at time $k$, $x_k$ to the state at time $k$, $u_k$ to the corresponding, cost minimizing control input at time $k$ while $\mathcal{U}$ is the (allowed) solution space for the control inputs $u_k$ and $\mathcal{X}$ the respective (allowed) solution space for the system states $x_k$. \\

\subsection{Model Predictive Frequency Control}
\label{sec:mpfc}

The model predictive frequency controller uses the directly available system output $\Delta f$ as control signal. The control input (determined by the controller) is the regulation power $P_{\textrm{add}}$ that is supplied to the power system (e.g. by a battery).

The system, implemented within the MPC controller, based on the model in equation \ref{eq:linsys} derived in section \ref{sec:modelling}, is the discretization of:
\begin{align}
 \dot{x} &= A x + B u \nonumber\\
 \dot{x} &= \begin{bmatrix} \frac{-f_0}{2HS_{B}D_{l}} & 0 \\ 0 & \frac{-v}{C_{\textrm{bat}}}\end{bmatrix} x + \begin{bmatrix}\frac{f_0}{2HS_{B}} \\ \frac{-1}{C_{\textrm{bat}}} \end{bmatrix} u \label{eq:cont_model} 
\end{align}

with
\be x= \begin{bmatrix} \Delta f \\x_{\textrm{SoC}} \end{bmatrix} \textnormal{  and  } u=P_{\textrm{add}}.  \nonumber \ee

The first line in equation \ref{eq:cont_model} refers to the swing equation, while the second line captures the state of charge by integrating and accounts for losses. The MPC cost function weights\footnote{The ramp rate $\delta u_k$ is not penalized, as this not of substantial interest for energy storages like batteries, that can ramp power output very fast.} are defined as $\mathcal{Q}$, $\mathcal{R} > 0$, with $N = 2...50$ steps and a sampling time of $T_s = 100$ ms:

\begin{eqnarray}
\min_{u} & & \sum_{k=1} ^{N} \left[x_{k}^T\mathcal{Q}x_{k}+u_{k}^T\mathcal{R}u_{k}\right] \nonumber\\
\mbox{s.t.} & x_0 & = x(0) \nonumber\\
 			& x(k+1) & = Ax(k) + Bu(k)\nonumber\\
			& x_{min} & \leq x(k) \leq x_{max}\nonumber\\
			& u_{min} & \leq u(k) \leq u_{max}\nonumber\\
			& \delta u_{min} & \leq u(k)-u(k-1) \leq \delta u_{max} \label{eq:MPFC_QP_problem}.
\end{eqnarray}

%
%
%
%

\section{Employing Stability Constraints}
\label{sec:guaranteedstability}

\subsection{Control Lyapunov Function}
\label{sec:CLF}

Derived from Lyapunov theory, another approach for guaranteeing stability can be found. The new approach is based on a finite horizon optimal control problem with terminal cost. By inserting a so-called \emph{Control Lyapunov Function} (CLF) as a terminal cost term

\be \varphi(x(t+T_p)) := V(x(t+T_p))\nonumber \ee

into a generalized MPC setup, stability is achieved.

For a nonlinear system 
\be \dot{x}=f(x,u),\nonumber \ee
where $x \in \mathbb{R}^n$, $u \in \mathbb{R}^m$ and 
\be \dot{V}(x,u) = V_x\cdot f(x,u)\nonumber \ee 
a CLF is a proper, positive definite function $V:\mathbb{R}^n \to \mathbb{R}_+$ such that: 
\be \inf_u \left[ \dot{V}(x,u)\right] \leq 0.\nonumber \ee 

If it is possible to ensure a negative derivative at every point by an appropriate choice of $u$, the system can be stabilized with $V$.
It can be shown, that such a CLF $V$ exists, when a globally asymptotically stabilizing control law $u=k(x)$ exists (smooth everyhwere except at $x=0$, cf. \cite{sontag1989universal}). However, there do not exist systematic approaches to find suitable CLFs for given nonlinear systems \cite{clf_jadbabaie}. The terminal region is not restricted $W := \infty$.

\subsubsection{CLF based MPC}
\label{sec:clf_mpc}

To implement a Control Lyapunov Function as a stability guarantee into an MPC setup, the corresponding terminal cost has to be derived. For that, a quadratic penalization of the form is used: $x^T \mathcal{Q}_{\textrm{term}} x$, with $\mathcal{Q}_{\textrm{term}}$ being the solution $X$ of the Lyapunov equation: 
\be AXA^T-X+\mathcal{Q}=0. \label{eq:lyapunov} \ee
The solution $X$ is symmetric and positive definite, as $\mathcal{Q}$ is symmetric, positive definite and $A$ has all its eigenvalues inside the unit disk (time discrete case). This approach assumes no control action after the end of the horizon, so that
\be x(k+i+1) = Ax(k+i), i=N,...,\infty. \nonumber \ee
This only makes sense if the system is asymptotically stable, or no solution will exist.

\subsection{Passivity}
\label{sec:passivity}

As with the Zero Terminal State constraint and the Control Lyapunov Functions described in the preceding part, closed-loop stability can also be guaranteed by imposing passivity for a model predictive control problem. In \cite{primbsdoyle} a nonlinear model predictive control scheme was proposed, which is based on the optimal control, nonlinear model predictive control and Control Lyapunov Function. Analogously in \cite{passivity_raff} a nonlinear model predictive control scheme was developed, which makes use of the relationship between passivity and optimal control, as well as the relationship between nonlinear model predictive control and optimal control. \\
In the sequel, an introduction into the concept of passivity is given. Passivity is tied to optimal control by an input affine system, which is optimal if and only if it satisfies a passivity property with respect to the optimal feedback \cite{sepulchre1997constructive}. 
Based on this, it is shown in \cite{passivity_raff}, that passivity and (nonlinear) model predictive control can be merged together, such that the individual advantages of each concept can be maintained, i.e. feasibility and closed-loop stability due to passivity as well as good performance due to online optimization in nonlinear model predictive control.

Following \cite{passivity_raff}, consider the following affine nonlinear system 
\begin{eqnarray}
\dot{x} & 	= & f\left(x\right)+g\left(x\right)u\nonumber\\
 	y 	&	= & h\left(x\right),\label{eq:affine_nonlinear}
\end{eqnarray}

where $x \in \mathbb{R}^n$ is the state, $u \in \mathbb{R}^m$ is the input and $y \in \mathbb{R}^m$ the output. Local Lipschitz continuity as well as $(x,u)=(0,0)$ being an equilibrium point is assumed. Then, system \ref{eq:affine_nonlinear} is said to be passive, if there exists a positive semidefinite storage function $S$, such that the following is satisfied for all $t_0 \geq t_1$:

\be S(x(t_1)) - S(x(t_0)) \leq \int_{-t_0}^{t_1} u^T(t)y(t)\;\mathrm{d}t, \label{eq:storage_function}\ee
where $(u(t),x(t),y(t))$ is a solution of \ref{eq:affine_nonlinear}. If $S$ is differentiable as a function of time, then the this relation can be reduced to
\be \dot{S}(x(t)) \leq u^T(t)y(t)\label{eq:storage_function_dot}.\ee

In case, system \ref{eq:affine_nonlinear} has a well-defined normal form, a further characterization of passive systems is also possible in terms of relative degree and minimum-phase property. Specifically, the relative degree needs to be $r=1$, i.e. $L_gh(0) \neq 0$, and the system must be \emph{weakly minimum-phase}.

\subsubsection{Passivity and Stability}

To achieve asymptotic stability of system \ref{eq:affine_nonlinear}, one can make use of the relationship between Lyapunov stability and the concept of passivity. It can be established by using the storage function $S(x)$ as Lyapunov function. As passivity only requires $S(x)$ to be \emph{positive semi-definite}, the equlibrium point $x=0$ might be unstable, even if passivity is ensured. This would be the case, if there exists an unobservable part of the system that is unstable. Hence, this unobservable part needs to be asymptotic stable for system \ref{eq:affine_nonlinear}, i.e. for its solution holds:
\be u = 0 \textnormal{ satisfies } \lim_{x \to \infty} x(t) = 0 \textnormal{ for } y(t) = 0 \textnormal{ and } t \geq 0, \label{eq:zero-state-detectable}\ee
which is is the so called property of \emph{zero-state detectability}.

In \cite{khalil} a concise summary of the key points describing the relationship between passivity and Lyapunov stability is given:

\begin{itemize}
\item The equilibrium point, $x = 0$, of the system \ref{eq:affine_nonlinear} with zero input, $u = 0$, is \emph{stable}, if the storage function $S(x)$ is \emph{positive semi-definite} and the system is \emph{passive} and \emph{zero-state detectable}.
\item The equilibrium point, $x = 0$, of the system \ref{eq:affine_nonlinear} with zero input, $u = 0$, is \emph{stable}, if the storage function $S(x)$ is \emph{positive definite} and the system is \emph{passive}.
\item The equilibrium point, $x = 0$, of the system \ref{eq:affine_nonlinear} with zero input, $u = 0$, is \emph{asymptotically stable}, if the storage function $S(x)$ is \emph{positive definite} and the system is \emph{strictly passive}.
\end{itemize}

In this case, a semidefinite storage function $S(x)$ is considered. Assuming zero-state detectability, the system can be stabilized with the feedback $u=-y$. With equation \ref{eq:storage_function_dot}, the following relation can be established:

\be \dot{S}(x(t)) \leq u^T(t)y(t) \leq -y^T(t) y(t) \leq 0,\label{eq:storage_passivity}\ee

and $S(x(t)) = 0 \to h(x(t)) = 0$ \cite{sepulchre1997constructive}. Furthermore, if $S(x)$ is radially unbounded, i.e. $S(x) \to \infty$ for $\|x\| \to \infty$, and all solutions are bounded, then the system \ref{eq:affine_nonlinear} can be globally stabilized by feedback $u=-y$.

Please note here, that \ref{eq:storage_passivity} can promptly be rewritten to an inequality, which easily adds to common MPC setups as a state constraint:

\be u^T(t) y(t) + y^T(t) y(t) \leq 0. \label{eq:passivity_constraint} \ee

Unfortunately, the derived passivity constraint is generally not convex, written for a time-discrete setup:
\be u^T(k) y(k) + y^T(k) y(k) \leq 0 \nonumber \ee 

with $y = x$ and written for the case with one state $x_1$ and one control input $u_1$:
\begin{align*}
 x_1u_1 + x_1^2 & = \begin{bmatrix} x_1 & u_1 \end{bmatrix}  
 \mathcal{M}
\begin{bmatrix} x_1 \\ u_1 \end{bmatrix} \nonumber \\
 				& = \begin{bmatrix} x_1 & u_1 \end{bmatrix}  
\begin{bmatrix} 1 & 1/2 \\ 1/2 & 0\end{bmatrix} 
\begin{bmatrix} x_1 \\ u_1 \end{bmatrix} \nonumber 
\end{align*}

where $\mathcal{M}$ has the eigenvalues $\lambda_1 \approx 1.21$ and $\lambda_2 \approx -0.21$, which means, that $\mathcal{M}$ is not positive (semi-)definite.
Actually, by a similar argument, $x(k)u(k) \leq 0$ is bilinear and therefore never convex.

To obtain a convex QP optimization problem, which can be solved by standard solvers, the passivity constraint is implemented in a way, such that it only holds for the first (applied) sample $k=1$ of the receding horizon. With this, $x(1)$ is fixed, a linear and hence convex constraint is obtained:
\be u^T(1) x(1) + x^T(1) x(1) \leq 0. \label{eq:passivity_convex} \ee 

\paragraph{Passivity and Optimality}

In addition to the statements given in the sections before, the relationship between optimal control and passivity can be established by using the value function $V^*$ as a storage function $S$ and the otpimal feedback $u^*$ as an output of the system \ref{eq:affine_nonlinear}. Taking the optimal feedback, which stabilizes the considered system, it minimizes the performance index $V^*$ if and only if the system

\begin{eqnarray}
\dot{x} & 	= & f\left(x\right)+g\left(x\right)u\nonumber\\
 	y 	&	= & k^*\left(x\right),\label{eq:affine_nonlinear_augm}
\end{eqnarray}

is zero-state detectable and output feedback passive with respect to 

\be \dot{S}(x(t)) \leq u^T(t)y^*(t) + \frac{1}{2} y^{*T}(t) y^*(t),\nonumber\ee

with $S = \frac{1}{2} V^*,$ following \cite{passivity_raff}. For a more rigerous argument, please be referred to \cite{sepulchre1997constructive}.

\subsubsection{Passivity based MPC}
\label{sec:passivity_mpc}

The in the preceding parts of this article, both (nonlinear) model predictive control and passivity were presented. Here, the aim is to merge these two concepts, inspired by the relationships between optimal control and passivity and between optimal control and nonlinear model predictive control \cite{primbsdoyle}.\\
The goal of merging those concepts is to maintain their corresponding individual advantages, i.e. good control performance due to on-line optimization within the nonlinear model predictive scheme and guaranteed closed-loop stability due to passivity, referred to as \emph{passivity based NMPC}.

Suppose a general NMPC setup, with a passivity constraint added (last line):

\begin{eqnarray}
 \min\limits_{u} 	& & \int_{t}^{t+T_p} \left(q(x(\tau))+u^T(\tau)u(\tau)\right)d\tau \nonumber \\
	s.t. 	& &\dot{x} = f\left(x\right)+g\left(x\right)u\nonumber \\
 			& & y  = h(x) \nonumber \\
			& & u^T(t) y(t) + y^T(t) y(t) \leq 0. \label{eq:passivity_ocproblem}
\end{eqnarray}

Where the last line directly follows from \ref{eq:passivity_constraint} as a passivity based constraint. In case the system \ref{eq:affine_nonlinear} is passive and zero-state detectable, it can be stabilized with the feedback $u=-y$ (cf. \cite{primbsdoyle}). Therefore, the passivity based constraint is a stability constraint which guarantees closed-loop stability. This holds even globally, if the storage function $S(x)$ is radially unbounded and all solutions of the system are bounded.\\
Unlike to many other nonlinear model predictive control schemes, which achieve stability by enforcing a decrease of the Control Lyapunov Function (CLF), $V(x)$, along the trajectory of the solution, stability is achieved directly with the aforementioned state constraint.

%

On first sight, the proposed scheme seems to be rather restrictive, as it is only applicable to passive systems. However, for stabilizing purposes, no real physical output $y$ is needed. It is enough to have a fictitious output $\eta = \sigma (x)$, yielding the following (fictitious), passive system:
\begin{eqnarray}
\dot{x} & 	= & f\left(x\right)+g\left(x\right)u\nonumber\\
 	\eta &	= & \sigma\left(x\right).\label{eq:passive_fict}
\end{eqnarray}
Note, that there always exists such a fictitious output $\eta$, if a Control Lyapunov Function exists, since then, by definition $\frac{\partial V}{\partial x}g(x)$ is a (fictitious) passive output. Unfortunately, there is no standardized way to construct a passive output. But as passivity is a physically inherited concept, such a fictitious passive output can often be found.\\
%
%
%

\paragraph{Applied Passivity Constraint}
In the following, the applied passivity constraint will be derived. As stability with respect to the frequency deviations $x_1 = \Delta f_1$ and $x_2 = \Delta f_2$ should be secured, a storage function of the form is taken:
\be S(x) = \frac{1}{2} \beta_1 x_1^2 + \frac{1}{2} \beta_2 x_2^2. \label{eq:storage_function2} \ee

The goal is, to fulfill the passivity property
\be u^T(t) y(t) + y^T(t) y(t) \leq 0 \label{eq:passivity_property2} \ee
with 
\be \dot{S}(x) \leq u^T(t) y(t).\label{eq:passivity_property} \ee

First, the constant $\beta_i$ is derived, which is a proportionality factor of the frequency $f$, describing the kinetic energy $E_{kin,i}$ stored in the rotating machines of area $i$ around the setpoint $f \approx f_0$: 
\be E_{kin,i} = H_i S_B = \frac{1}{2} J_i \omega_m^2 = \frac{1}{2} J_i (2 \pi)^2 f^2 = \frac{1}{2} \beta_i f^2 \approx \frac{1}{2} \beta_i f_0^2, \nonumber \ee
with 
\be \beta_i = J_i (2 \pi)^2 = \frac{2 H_i S_{B,i}}{f_0^2} \geq 0.\label{eq:beta} \ee 
Assuming two machines or areas $i=\{1,2\}$, the storage function from equation \ref{eq:storage_function2} becomes: 
 \begin{align}
\dot{S} & = \beta_1 x_1 \dot{x}_1 + \beta_2 x_2 \dot{x}_2   \nonumber \\
 & = \beta_1x_1(A_{\textrm{freq},1}x_1 + B_{\textrm{freq},1}u_1) + \beta_2x_2(A_{\textrm{freq},2}x_2 + B_{\textrm{freq},2}u_2) \label{eq:storage_function_expanded}, 
\end{align}

where  $ \beta_iA_{\textrm{freq},i}x_i^2 \leq 0 \textnormal{ for $
 i=\{1,2\}$}$
is true for all $x_i \in \mathbb{R}$, since $\beta_i \geq 0$ and 
$ A_{\textrm{freq,}i} = \frac{-f_0}{2H_i S_{B,i}D_{l,i}} \leq 0 $.\\
The passivity property $ \dot{S} \leq u^T y = u_1 y_1 + u_2 y_2 $ (cf. equation \ref{eq:passivity_property})  hence yields
\begin{align}
\dot{S} & =  \underbrace{\beta_1x_1A_{\textrm{freq},1}x_1}_{\leq 0} + \beta_1x_1B_{\textrm{freq},1}u_1 + \underbrace{\beta_2x_2A_{\textrm{freq},2}x_2}_{\leq 0} + \beta_2x_2B_{\textrm{freq},2}u_2 \nonumber\\
& \leq  u_1y_1 + u_2y_2. \label{eq:storage_function_expanded2} 
\end{align}
Combined 
with equation \ref{eq:beta}, $B_{\textrm{freq},i} = \frac{f_0}{2H_i S_{B,i}} = \frac{1}{\beta_i}\cdot\frac{1}{f_0} $ and as well neglecting all (marked) negative terms, equation \ref{eq:storage_function_expanded2} can be transformed into 

\be u_1 \left(\frac{x_1}{f_0}\right) + u_2 \left(\frac{x_2}{f_0}\right) \leq u_1y_1 + u_2y_2.\label{eq:passivity_condition} \ee

Taking $y_i = \beta_i B_{\textrm{freq},i} x_i$ yields
$ y_i = \frac{x_i}{f_0} $ and together with the passivity property $u^T(t) y(t) + y^T(t) y(t) \leq 0 $ (cf. equation \ref{eq:passivity_property2}) finally results in
\begin{align}
 u_1 y_1 + y_1^2 + u_2 y_2 + y_2^2 & \leq 0 \nonumber\\
\Leftrightarrow \quad u_{1}\left(\frac{x_{1}}{f_0}\right)+\left(\frac{x_{1}}{f_0}\right)^2 + u_{2}\left(\frac{x_{2}}{f_0}\right) + \left(\frac{x_{2}}{f_0}\right)^2 &\leq 0  \label{eq:passivity_condition2}
\end{align}

as the corresponding passivity constraint for a two-machine system. Analogously for the  one-machine system:
\begin{align}
 u y + y^2 &\leq 0 \nonumber\\
 \Leftrightarrow \quad   u \left(\frac{x_{1}}{f_0}\right)+\left(\frac{x_{1}}{f_0}\right)^2 &\leq 0. \label{eq:passivity_condition2_onesys}
\end{align}

In the here presented implementation, the passivity based constraint is enforced only during the first sampling interval $t$ and not for the whole length of the finite prediction horizon $T_p$. This allows the same stability guarantees as enforcement of the constraint for the whole horizon does, while it generally exhibits better feasibility and less computational effort (cf. section \ref{sec:passivity}).

\subsection{Summary}

The passivity based nonlinear model predictive control approach combines passivity with model predictive control by demanding the derivative of a storage function $S(x)$ to fulfill $\dot{S}(x) \leq u^T y$ (being positive semi-definite). At the same time their respective advantages are maintained: Feasibility and closed-loop stability due to \emph{passivity} and good performance due to on-line optimization by \emph{model predictive control}. \\
A stability guaranteeing Control Lyapunov function ensures a storage function $V(x)$ to satisfy $\dot{V}(x) \leq 0$ (which actually needs to be positive definite). Where such a Lyapunov function cannot be found, the passivity based approach is especially appealing. 
The corresponding basic principles are the relationships of passivity and nonlinear model predictive control to optimal control.

\section{Study Case and Results}
\label{sec:study_case_results}


\subsection{General Setup}
\label{sec:general_setup}

The application analyzed for the derived stability concepts is to guarantee frequency stability in a power system. 
In this article, the focus lies on the evaluation of the interaction between two areas, interconnected with a tie-line. For this case the used model was derived in section \ref{sec:two_machine_model}.\\
All examples are modeled as a quadratic program (QP) and in per-unit system (p.u.) for scalability. The primary optimization objective is to maintain a frequency deviation $\Delta f = 0$, meaning, that the frequency should be close to $f_0 = 50$ Hz in every area. Internally, the frequency related states of the used model are normed to $x_i = \frac{\Delta f_i}{f_0}$. For illustrative purposes, in all shown diagrams, frequency related states are transformed to frequencies, hence they are shown with the unit Hz. 

For simulation and optimization purposes, the YALMIP environment was used, developed and maintained by Johan L\"ofberg. The software package used for mathematical optimization within the predictive control framework of YALMIP is CPLEX.
All simulations were run on a computer with a quad-core CPU (Intel Core i7, 3.4 GHz), with 16 GB memory and 64-bit Windows 7 as operating system.

\subsection{Developed Simulink Model}
\label{sec:simulink_model}
A model was developed within Simulink, which is part of a MATLAB package. 

\begin{figure}
	\centering
	\includegraphics[trim = 0cm 7cm 0.4cm 3cm, clip, width=0.9\textwidth]{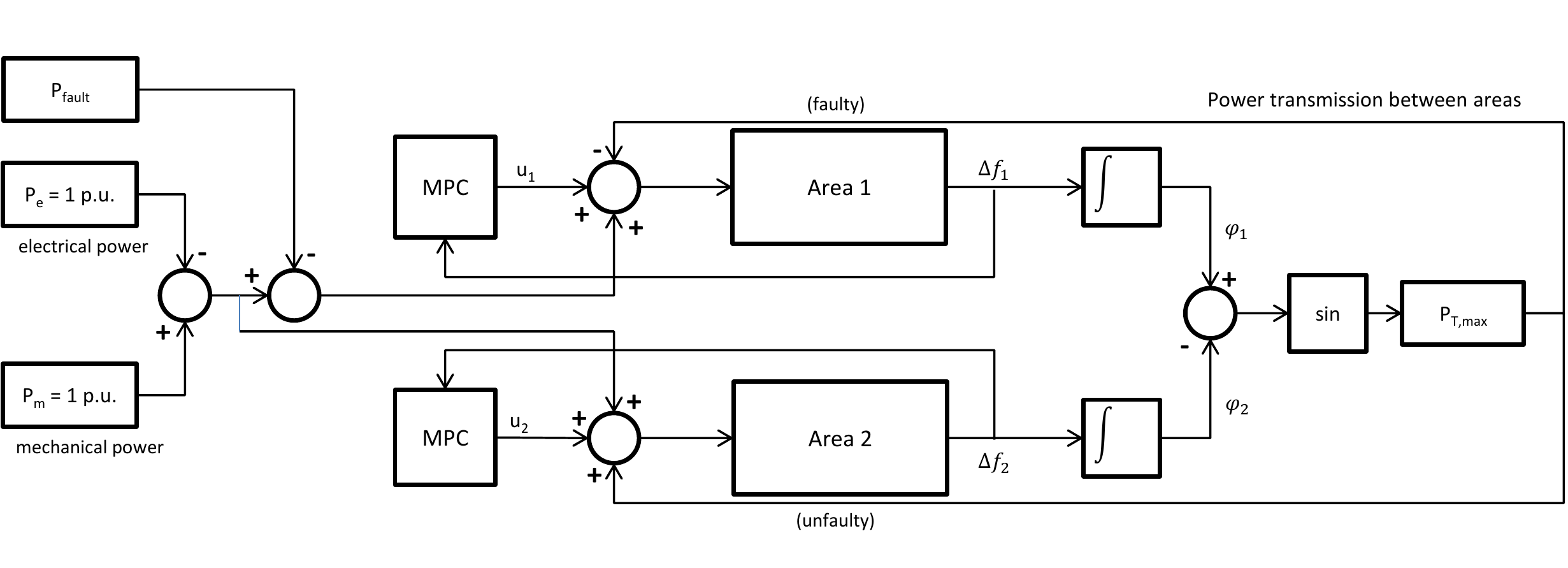}
	\caption{Overview of the implemented control scheme as an example for controlling a decentralized two area power system. Area 1 is faulty, which means, that a disturbing power $P_{fault}$ is applied to the expected balance of $P_e = P_m = 1$ p.u. for area 1. Area 2, however, is unfaulty, but connected to area 1 enabling power transmission between both areas.}
	\label{pic:control_scheme}
\end{figure}

An overview of the devloped Simulink model is provided in figure \ref{pic:control_scheme}.

\subsubsection{One Area}
For carrying out experiments in a one area setup, the control input for the second plant was disconnected and a zero transmission power ($P_{T,max}=0$) was assumed. The model equations used are time discretized versions of the derived model in section \ref{sec:mathematical_modeling}. The following state vector and input was used:
$ x = \begin{bmatrix}
\frac{\Delta f_{}}{f_0} & x_{\textrm{SoC}} \end{bmatrix}^T, $
$ u = \begin{bmatrix} u \end{bmatrix}. $

The MPC cost function weights\footnote{Please note at this point, that internally the frequency state penalized is normed to $\frac{\Delta f}{f_0}$.} are defined as:
\be \mathcal{Q} = \begin{bmatrix} 10 & 0 \\ 0 & 0.001 \end{bmatrix},\quad \mathcal{R} = 1, \nonumber \ee

with a prediction horizon length of $N = 2...50$ steps and a sampling time of $T_s = 100 \textnormal{ ms} \mathrel{\widehat{=}} 0.1$ s. Observe here that penalizing $u$ with $\mathcal{R} > 0$ has physical and/or economical reasons, but might lead to instabilities using the standard MPC setup. The following setup (derived in section \ref{sec:mpfc}) is referred to as the standard MPC setup:

\begin{eqnarray}
\min_{u} & & \sum_{k=1} ^{N} \left[x_{k}^T\mathcal{Q}x_{k}+u_{k}^T\mathcal{R}u_{k}\right] \nonumber\\
\mbox{s.t.} & x_0 & = x(0) \nonumber\\
 			& x(k+1) & = Ax(k) + Bu(k)\nonumber\\
			& x_{min} & \leq x(k) \leq x_{max}\nonumber\\
			& u_{min} & \leq u(k) \leq u_{max}\nonumber\\
			& \delta u_{min} & \leq u(k)-u(k-1) \leq \delta u_{max} \label{eq:MPFC_QP_problem2}.
\end{eqnarray}

For ensuring wide feasibility of the optimization algorithm, the limit of the frequency deviation was set to a rather big range (higher deviations might cause serious system damages): $\Delta f \in \left[-1.5\textnormal{ Hz},1.5\textnormal{ Hz}\right].$ The state of charge is limited to: $x_{\textrm{SoC}} \in \left[-0.75,0.75\right].$ Initially, both states are assumed to equal zero.
The battery power is constrained with \be u_{\textrm{MPC}} \in \left[-0.15 \textnormal{ p.u.},0.15 \textnormal{ p.u.}\right].\nonumber\ee
A rate constraint is implemented as well. It is assumed, that a battery can change the power by 1 p.u. at each sample, hence \be\delta u = u_k - u_{k-1} \in \left[\frac{-1}{T_s}, \frac{1}{T_s}\right].\nonumber\ee

%
%
%
%

\paragraph{Setting up Passivity}

To ensure passivity, the following constraint was employed to the first (applied) sample of the optimized control input (cf. equations \ref{eq:passivity_convex} and \ref{eq:passivity_condition2_onesys}):
\be u_{\textrm{MPC}} \left(\frac{\Delta f_{}}{f_0}\right)+\left(\frac{\Delta f_{}}{f_0}\right)^2 \leq 0, \nonumber\ee
hence, applied for the first sample: \be u_{\textrm{MPC}}(1) \cdot x_{1}(1)+x_{1}(1)^2 \leq 0. \nonumber\ee

\paragraph{Setting up the CLF}
\label{sec:onearea_clf}

To setup the CLF based MPC, the terminal costs for the one area system are defined as the solution of the (discrete) Lyapunov equation\footnote{In Matlab: \lstinline$q_term = dlyap(A(1,1),Q(1,1))$.}, which results in $q_{\textrm{term}} = 40005.$
This is one dimensional, as it only refers to stability of the first state, the frequency deviation. Stability of the second state, the state of charge, is implied by constraining the integrated control action. Therefore, the corresponding terminal costs for the second state are set to zero:
\be \mathcal{Q}_{\textrm{term}} = \begin{bmatrix} 40005 &  0 \\ 0 & 0 \end{bmatrix}.\nonumber\ee


\subsubsection{Two Areas (Uncoordinated Case)}
The uncoordinated two area system consists basically of two equal one area systems (each one comprising the same MPC setup as in the one area case), where power transmission between the two systems is activated. Both systems are controlled individually with an own energy storage system, therefore, all parameters and constraints are set as in the one area system mentioned before.\\
The power transmission is implemented as derived in section \ref{sec:two_machine_model} in the form of:
\be P_{T12} = \hat{P}_T\sin(\varphi_1 - \varphi_2), \nonumber \ee
 with a maximum power transmission of $\hat{P}_T = 0.2 \textnormal{ p.u.}$

\paragraph{Setting up Passivity}
Passivity is set as a constraint for each system individually:
\be u_{\textrm{MPC},1} \left(\frac{\Delta f_{1}}{f_0}\right)+\left(\frac{\Delta f_{1}}{f_0}\right)^2 \leq 0, \nonumber\ee
\be u_{\textrm{MPC},2} \left(\frac{\Delta f_{2}}{f_0}\right)+\left(\frac{\Delta f_{2}}{f_0}\right)^2 \leq 0, \nonumber\ee

with $x_1 := \left(\frac{\Delta f_{1}}{f_0}\right)$ and $x_2 := \left(\frac{\Delta f_{2}}{f_0}\right)$ it follows applied for the first sample:

\be u_{\textrm{MPC}}(1) \cdot x_{1}(1)+x_{1}(1)^2 \leq 0, \nonumber\ee 
\be u_{\textrm{MPC}}(1) \cdot x_{2}(1)+x_{2}(1)^2 \leq 0. \nonumber\ee 

\paragraph{Setting up the CLF}

The CLF setup is done as in the one area control case (cf. section \ref{sec:onearea_clf}), individually for each controller.


\subsubsection{Two Areas (Coordinated Case)}
\label{sec:two_areas_coordinated_model}

In the coordinated controlled version of the two area system, the power angle between the two systems is introduced as a new system state. The corresponding model equations as well as the new state space matrices reflecting the coupling between the systems were derived in section \ref{sec:two_machine_model}. Still, it is assumed, that every area uses its own energy storage unit, hence the two area system consists of the state vector
$ x = \begin{bmatrix}
\frac{\Delta f_{1}}{f_0} & x_{\textrm{SoC},1} & \frac{\Delta f_{2}}{f_0} & x_{\textrm{SoC},2} & \Delta \varphi
\end{bmatrix}^T $
and input vector
$ u = \begin{bmatrix}
u_1 & u_2 \end{bmatrix}^T. $

The cost function weights hence change and were defined as the following for the coordinated case:

\be \mathcal{Q} = \begin{bmatrix} 
10 & 0 & 0 & 0 & 0 \\ 
0 & 0.001 & 0 & 0 & 0 \\
0 & 0 & 10 & 0 & 0 \\
0 & 0 & 0 & 0.001 & 0 \\
0 & 0 & 0 & 0 & 0.1\\ \end{bmatrix}, 
\quad \mathcal{R} = \begin{bmatrix} 1 & 0 \\ 0 & 1 \end{bmatrix}. \nonumber \ee

Besides that, all parameters and constraints are the same as in the uncoordinated two area system. 

\paragraph{Setting up Passivity}
The passivity constraint is adapted to account for both frequency related states $x_1$ and $x_2$ and the same time:
\be u_{\textrm{MPC},1}\left(\frac{x_{1}}{f_0}\right)+\left(\frac{x_{1}}{f_0}\right)^2 + u_{\textrm{MPC},2}\left(\frac{x_{2}}{f_0}\right)+\left(\frac{x_{2}}{f_0}\right)^2 \leq 0,\nonumber\ee
respectively:
\be u_{\textrm{MPC},1}(1) x_1(1)+x_1(1)^2 + u_{\textrm{MPC},2}(1)x_2(1)+x_2(1)^2 \leq 0.\nonumber\ee

\paragraph{Setting up the CLF}
For the coordinated version, CLF stability for both frequency deviations $\Delta f_1, \Delta f_2$ is desirable at the same time, hence the solution of the (discrete) Lyapunov equation becomes the terminal cost as in the following matrix, accounting for both frequency stability related states:\be q_{\textrm{term}} = 10000 \cdot \begin{bmatrix}
2.2137 &  1.7868 \\ 1.7868 & 2.2137 \end{bmatrix}.\nonumber\ee The corresponding terminal costs for the other states as the angle and the states of charge are set to zero:
 \be \mathcal{Q}_{\textrm{term}} = 10000 \cdot \begin{bmatrix} 
 2.2137 & 0 & 1.7868 & 0 & 0 \\ 
 0 & 0 & 0 & 0 & 0 \\
 1.7868 & 0 & 2.2137 & 0 & 0\\
 0 & 0 & 0 & 0 & 0 \\
 0 & 0 & 0 & 0 & 0 \\
 \end{bmatrix}. \nonumber \ee

\subsubsection{Summary}
To summarize, besides the standard MPC setup (as in equation \ref{eq:MPFC_QP_problem2}) the applied stability approaches are implemented as following (generalized for both one and two area control). 

\paragraph{Standard MPC Setup}
 \begin{eqnarray}
\min_{u} & & \sum_{k=1} ^{N} \left[x_{k}^T\mathcal{Q}x_{k}+u_{k}^T\mathcal{R}u_{k}\right] \nonumber\\
\mbox{s.t.} & x_0 & = x(0) \nonumber\\
 			& x(k+1) & = Ax(k) + Bu(k)\nonumber\\
			& x_{min} & \leq x(k) \leq x_{max}\nonumber\\
			& u_{min} & \leq u(k) \leq u_{max}\nonumber\\
			& \delta u_{min} & \leq u(k)-u(k-1) \leq \delta u_{max} \label{eq:MPFC_QP_problem_3}.
\end{eqnarray}
 
 \paragraph{Passivity based MPC Setup}
 As an additional constraint to system \ref{eq:MPFC_QP_problem_3}, the passivity constraint is incorporated:
\be u^T(1) y(1) + y^T(1) y(1) \leq 0. \nonumber\ee
 
 \paragraph{CLF based MPC Setup}
 Here, the terminal costs are adjusted, i.e. compared to the standard setup in system \ref{eq:MPFC_QP_problem_3}, the cost function changes to 
\be \min_{u} \quad \sum_{k=1} ^{N-1} \left[x_{k}^T\mathcal{Q}x_{k}+u_{k}^T\mathcal{R}u_{k}\right] + x_{N}^T\mathcal{Q}_{\textrm{term}}x_{N} \nonumber\ee 
\subsection{Results}
\label{sec:results}

Two different setups were analyzed. First, the uncoordinated case, where a decentralized control scheme was applied. Second, the coordinated case, where a centralized controller has knowledge about all (potentially distributed) system states.

First, the angle differences between the two systems are presented for a fixed prediction horizon $N$. Afterwards, the simulations were generalized for various different prediction horizons $N=2...50$. \\
The applied fault signal is the one shown in figure \ref{pic:Faulty_Chirp} for every simulation carried out. Note, that the error is not zero mean and becomes constantly zero at $t=60$ s.

\subsubsection{Two Areas (Uncoordinated Case)}
\begin{figure}
\begin{minipage}[hbt]{0.45\textwidth}
        \centering
	\includegraphics[trim = 1.9cm 7.1cm 1cm 8.75cm, clip, width=1\textwidth]{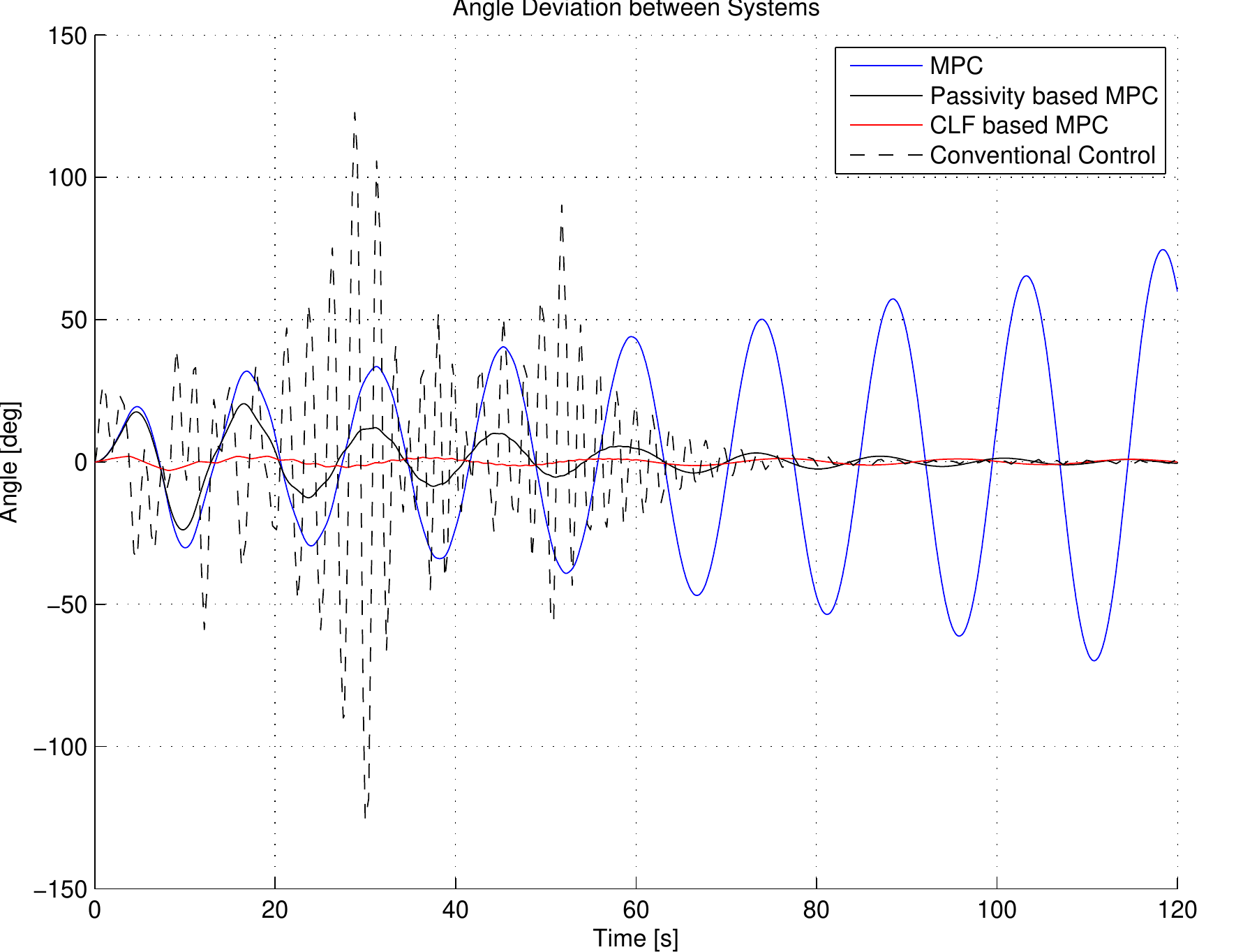}
	\caption{Angle differences between areas (uncoordinated), fixed $N = 3$.}
	\label{pic:AngleDeviation_N3_combined_2sys_cc}
\end{minipage}
\hfill
\begin{minipage}[hbt]{0.45\textwidth}
        \centering
	\includegraphics[trim = 1.9cm 7cm 1cm 8.7cm, clip, width=1\textwidth]{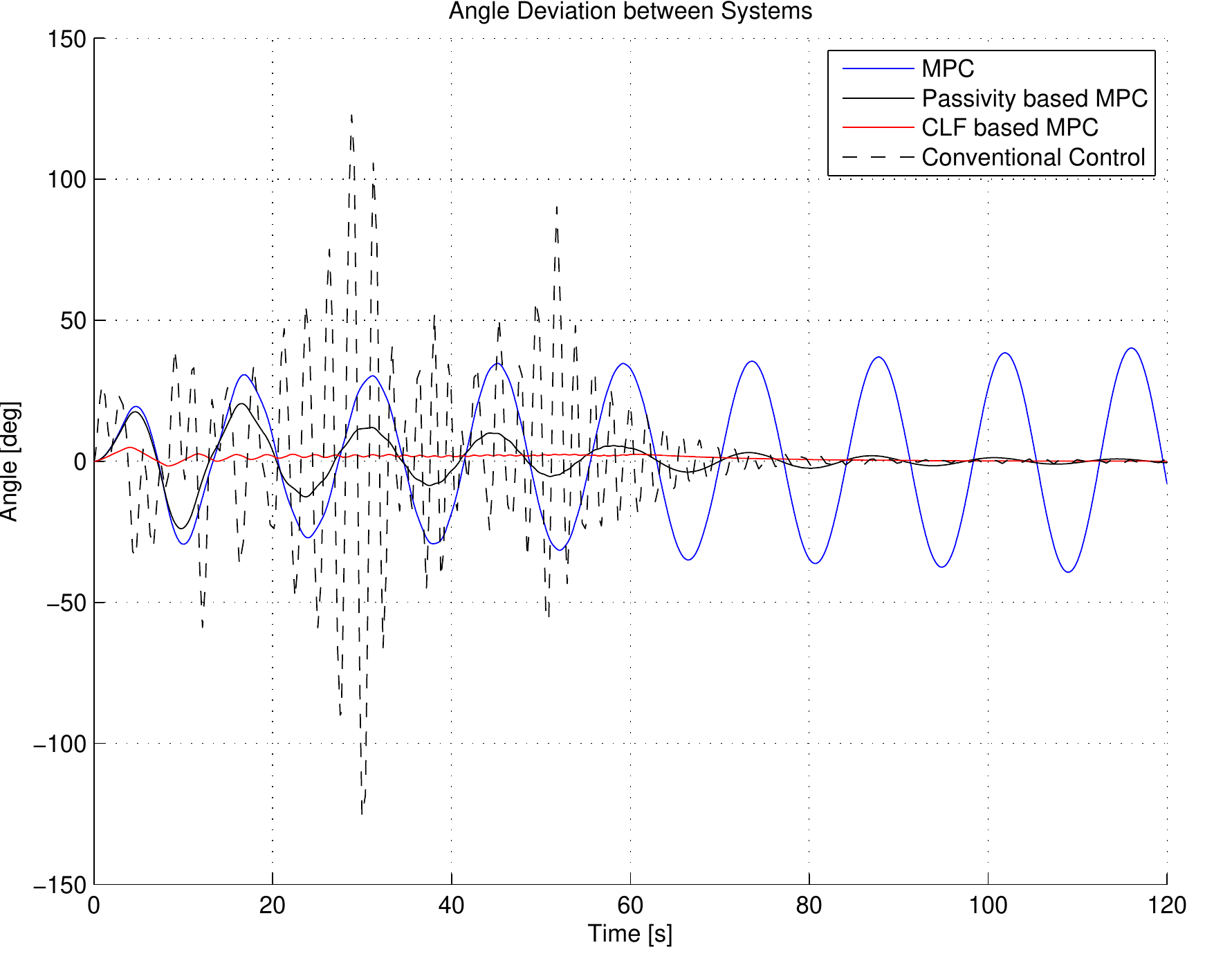}
	\caption{Angle differences between areas (coordinated), fixed $N = 3$.}
	\label{pic:AngleDeviation_N3_combined_2sys_coordinated_cc}
\end{minipage}
\end{figure}

Employing frequency stability for two equally sized, interconnected areas of power systems with a maximum transmission capacity of $\hat{P}_{T,max}=0.2$ p.u. was simulated.\\
The uncoordinated version was analyzed with respect to standard MPC control, passivity based MPC control and CLF based MPC control. 
Uncoordinated means here, that each area is controlled in an individual way and each controller has only local information, hence, a decentralized control topology. This implies a significant model-plant-mismatch, as the possibility of power transmission is not incorporated within the MPC model.\\
The effect of stability can be seen especially illustrative in figure \ref{pic:AngleDeviation_N3_combined_2sys_cc}, which shows the angle difference for the two controlled zones with respect to the different control approaches. Large angle differences (as conventional control exhibits) as well as increasing oscillations (as in standard MPC) indicate frequency instability.

\subsubsection{Two Areas (Coordinated Case)}
The power system itself is not altered. However, the area coupling is incorporated into the MPC setup in the coordinated approach. By that, the model-plant mismatch is significantly reduced compared to the uncoordinated case. Coordinated control refers to a centralized control topology, where the controller has full information about all states in all control areas.
Compared with the uncoordinated case, the stability behavior of a standard MPC approach improves, yet still exhibiting undesired persisting frequency oscillations. Essentially, the stability behavior of the different approaches remain unchanged.

%
%
%

\subsubsection{Generalizing Two Area Control}

\begin{figure}
\begin{minipage}[hbt]{0.48\textwidth}
        \centering
	\includegraphics[trim = 1.9cm 7.15cm 1cm 8.7cm, clip, width=1\textwidth]{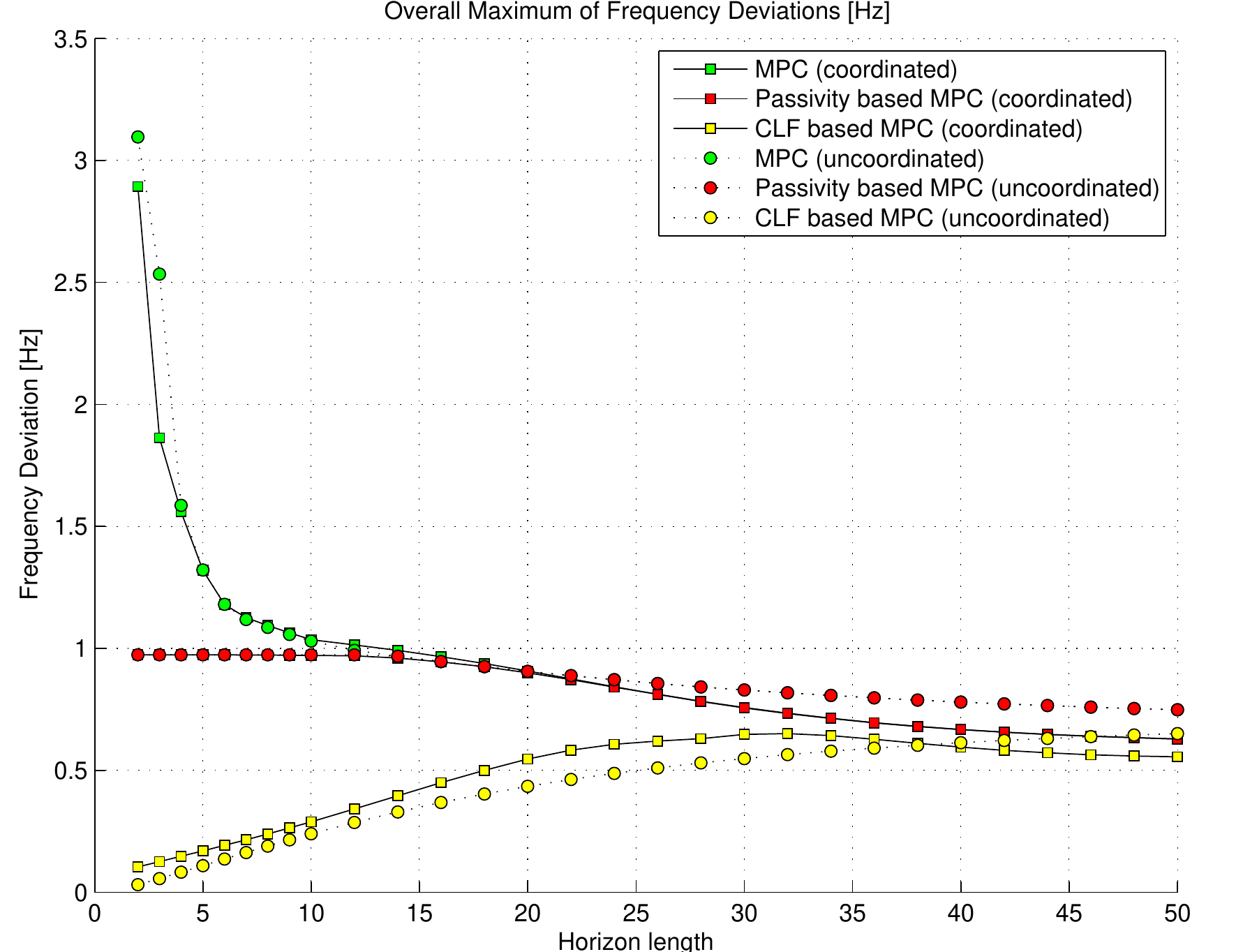}
	\caption{Maximum frequency deviations.}
	\label{pic:OverallMaxFreqDev_N_coupled_uncoupled}
\end{minipage}
\hfill
\begin{minipage}[hbt]{0.48\textwidth}
        \centering
	\includegraphics[trim = 1.9cm 7.15cm 1cm 8.7cm, clip, width=1\textwidth]{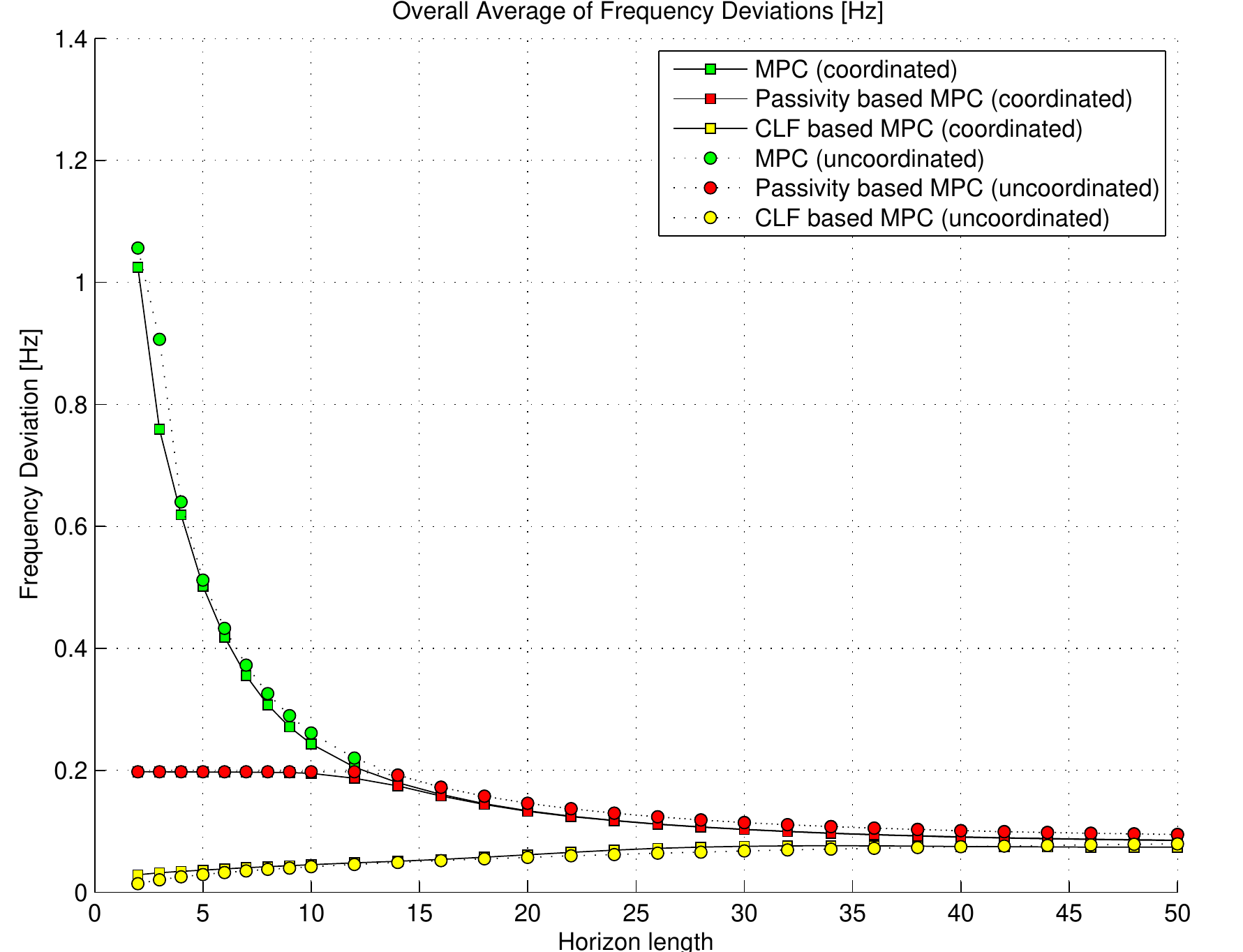}
	\caption{Average frequency deviations.}
	\label{pic:OverallAvgFreqDev_N_coupled_uncoupled}
\end{minipage}
\end{figure}

\begin{figure}
\begin{minipage}[hbt]{0.48\textwidth}
        \centering
	\includegraphics[trim = 1.9cm 7.15cm 1cm 8.7cm, clip, width=1\textwidth]{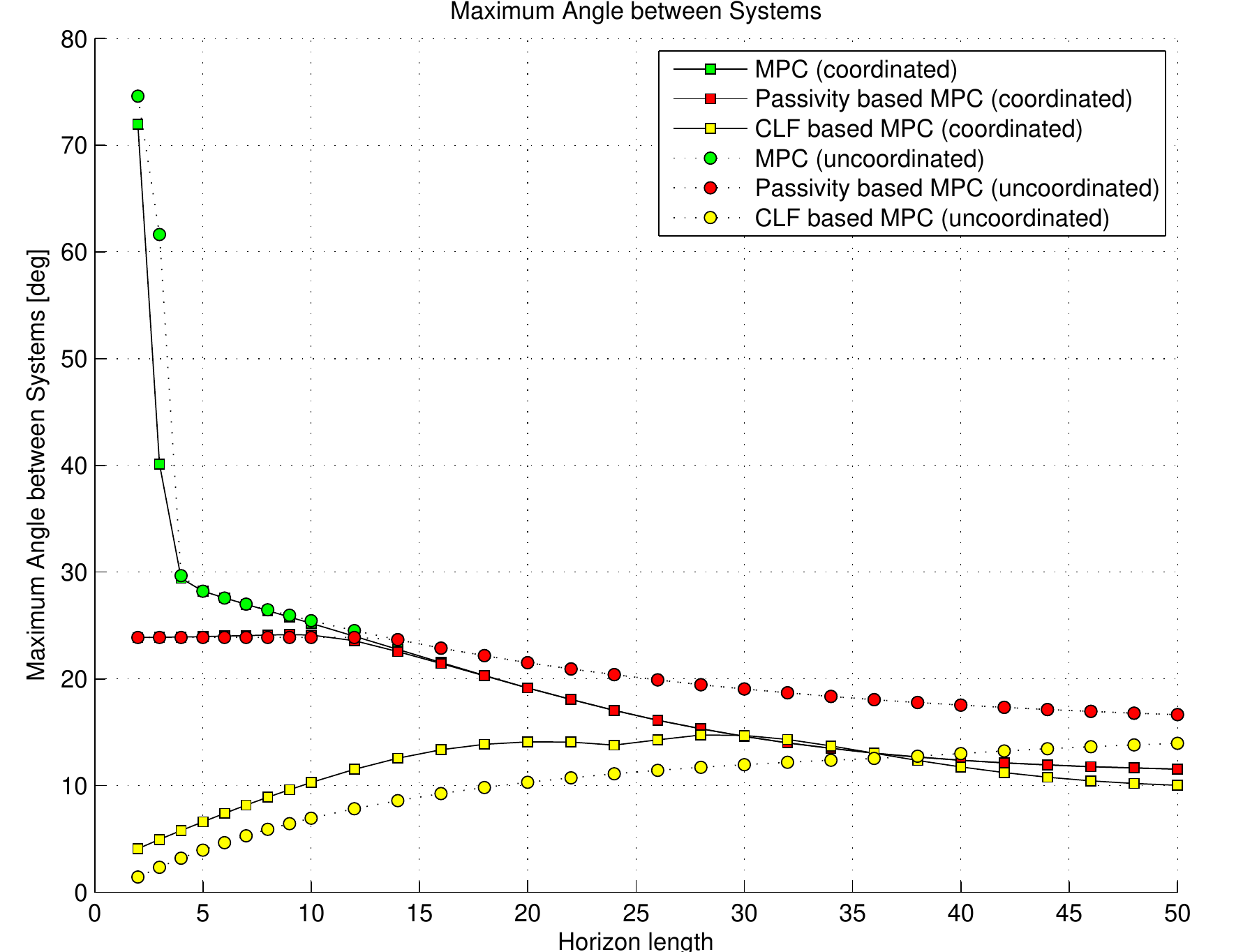}
	\caption{Maximum angle differences between areas.}
\label{pic:MaxAngleDeviation_N_coupled_uncoupled}
\end{minipage}
\hfill
\begin{minipage}[hbt]{0.48\textwidth}
        \centering
	\includegraphics[trim = 1.9cm 7.15cm 1cm 8.7cm, clip, width=1\textwidth]{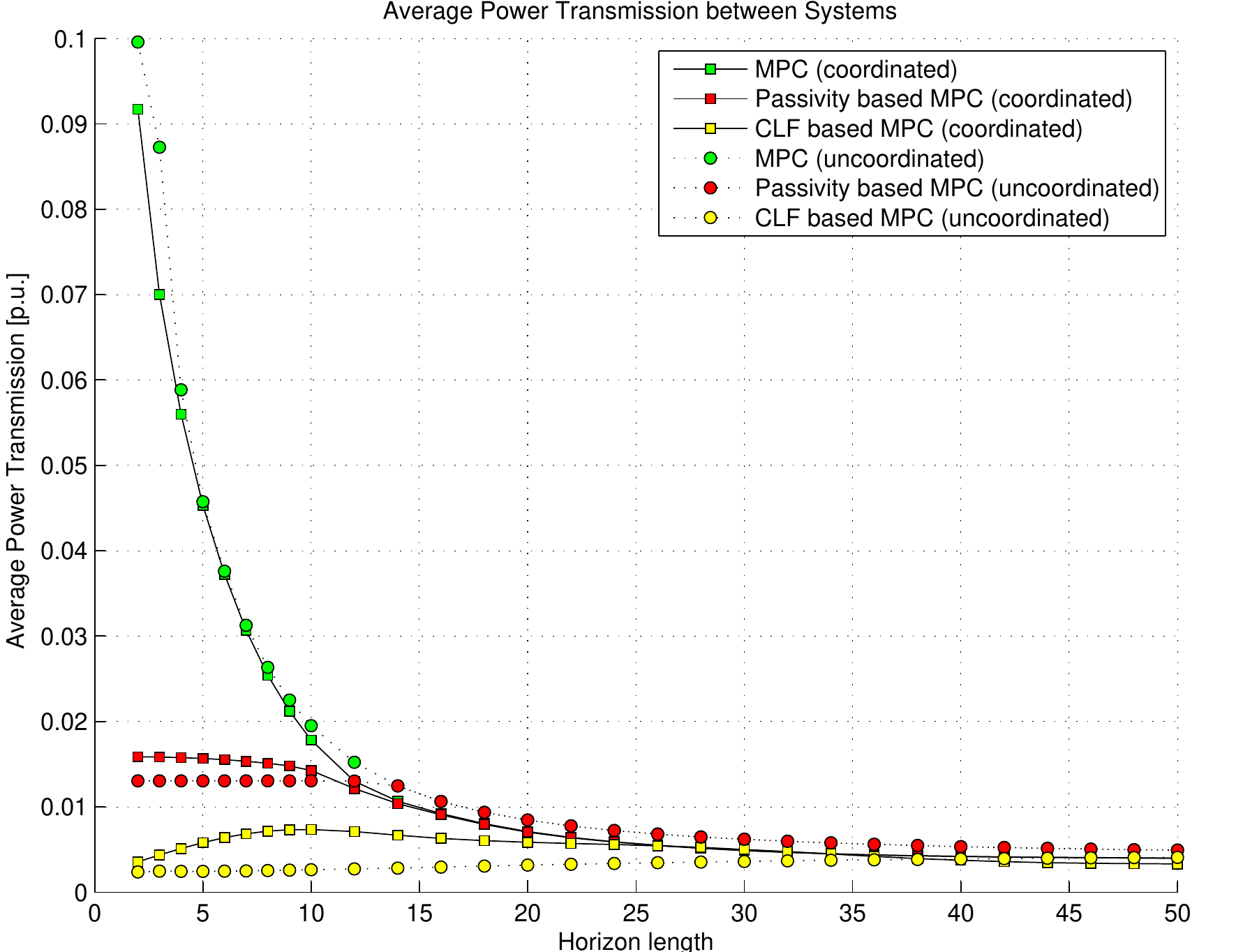}
	\caption{Average power transmission between areas.}
	\label{pic:AvgPowerTransmission_N_coupled_uncoupled}
\end{minipage}
\end{figure}

\begin{figure}
\begin{minipage}[hbt]{0.48\textwidth}
        \centering
	\includegraphics[trim = 1.8cm 7.15cm 1cm 8.7cm, clip, width=1\textwidth]{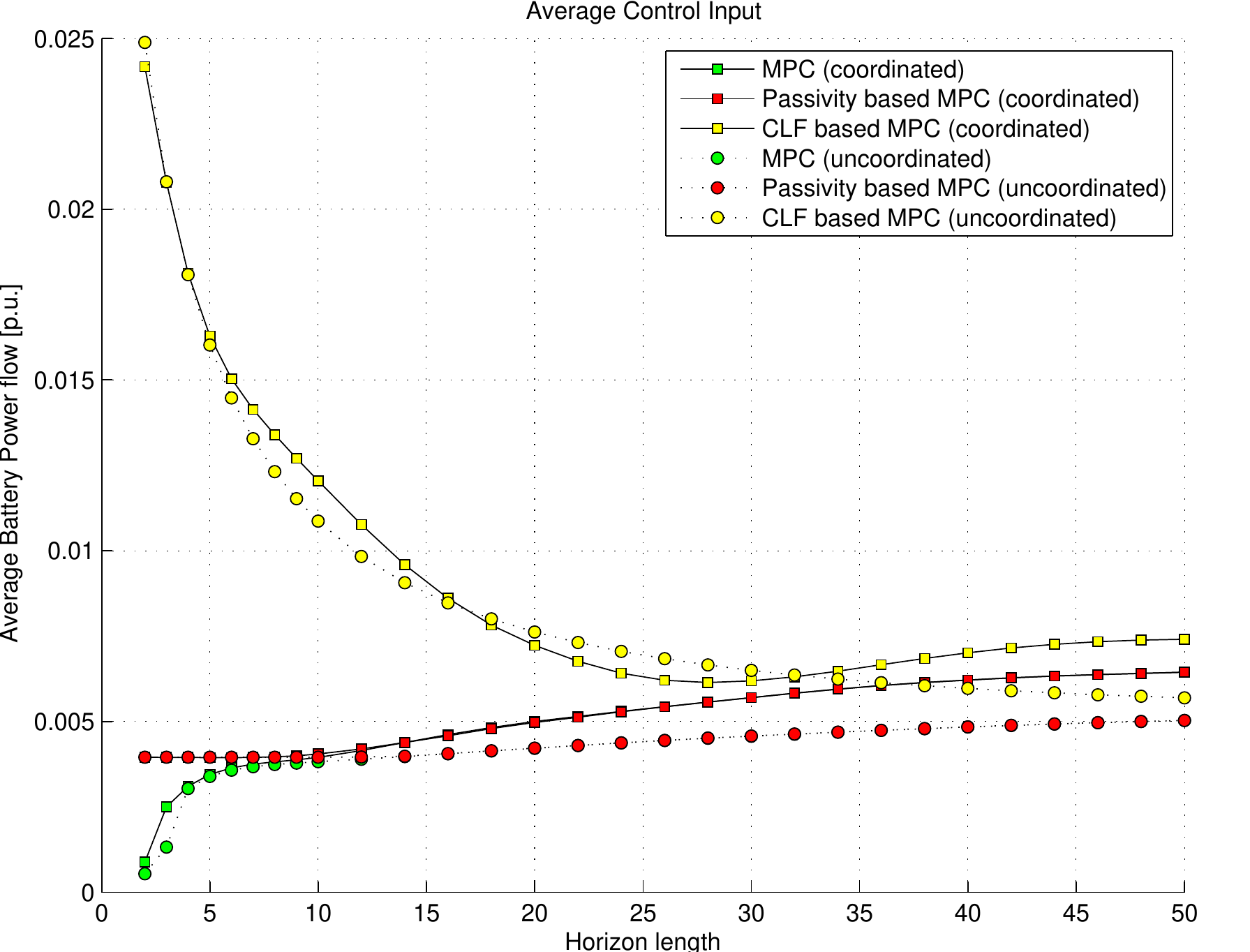}
	\caption{Average control input.}
	\label{pic:AvgBatteryPowerFlow_N_coupled_uncoupled}
\end{minipage}
\hfill
\begin{minipage}[hbt]{0.48\textwidth}
        \centering
	\includegraphics[trim = 1.8cm 7.15cm 1cm 8.7cm, clip, width=1\textwidth]{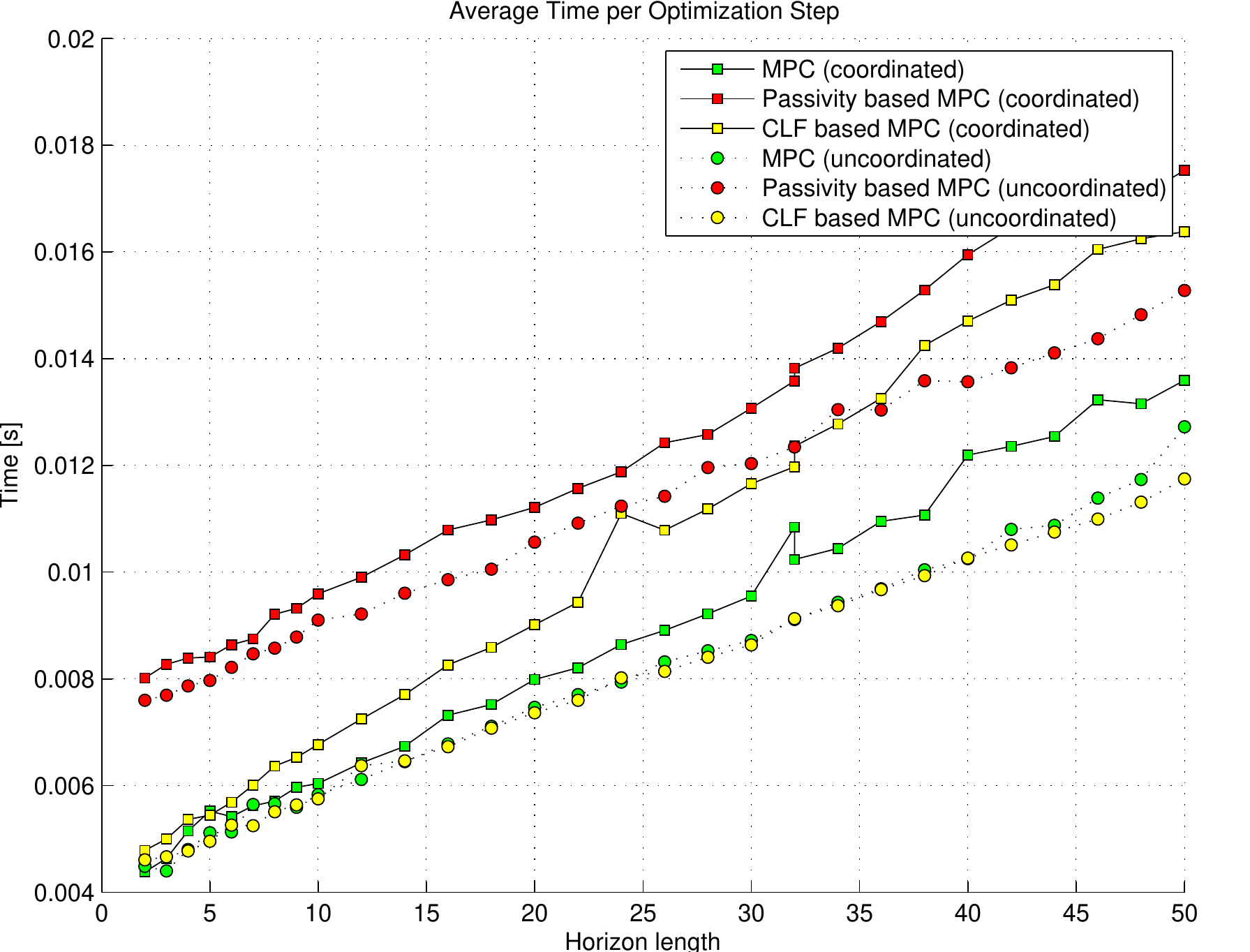}
	\caption{Time for each optimization step.}
	\label{pic:AvgTimeOptimIteration_N_coupled_uncoupled}
\end{minipage}
\end{figure}

The behavior of the states with respect to time was shown precedingly for a fixed prediction horizon $N=3$. Now, a generalization of controlling two areas for various prediction horizons $N=2...50$ is made.  
For facile comparisons, the performance measures of the presented MPC approaches with employed stability constraints for two uncoordinated areas are shown together with the performance measures for two coordinated areas in figures \ref{pic:OverallMaxFreqDev_N_coupled_uncoupled}, \ref{pic:OverallAvgFreqDev_N_coupled_uncoupled}, \ref{pic:MaxAngleDeviation_N_coupled_uncoupled}, \ref{pic:AvgPowerTransmission_N_coupled_uncoupled}, \ref{pic:AvgBatteryPowerFlow_N_coupled_uncoupled}, and \ref{pic:AvgTimeOptimIteration_N_coupled_uncoupled}. In these figure, control performance measures for two controlled systems (one being disturbed) over different horizon lengths, for two coordinated (continuous lines) and two uncoordinated areas (dashed lines) are compared.\\
Generally can be said, that the maximum frequency deviation for the coordinated and the uncoordinated version show rather similar characteristics (cf. figure \ref{pic:OverallMaxFreqDev_N_coupled_uncoupled}). In the case of standard MPC and passivity based MPC, the coordinated approach shows smaller frequency deviation than the uncoordinated approach, as expected and with inherent larger control input for coordinated version (cf. figure \ref{pic:AvgBatteryPowerFlow_N_coupled_uncoupled}). Interestingly, with a horizon length of less than 38 steps, the Lyapunov based MPC setup exhibits a smaller maximum frequency deviation for the uncoordinated case. If a prediction horizon between 40 and 50 steps is applied, this swaps and the coordinated case shows a smaller maximum frequency deviation.\\
Naturally, as can be seen in figure \ref{pic:AvgPowerTransmission_N_coupled_uncoupled}, a smaller frequency deviation causes less power transmission between the areas.\\
Referring to the time for each optimization step (which excludes the optimizer setup time), the situation is similar to the one area setup, where CLF based and standard MPC needs comparably less time than passivity based MPC. This behavior seems to be fairly independent of the control topology. The coordinated CLF based MPC approach appears to be an exception, as the slope looks considerably steeper than for all other MPC approaches.

\section{Conclusion and Outlook}
\label{sec:conclusion_outlook}

\subsection{Conclusion}
\label{sec:conclusion}

In this paper a model predictive control based frequency control scheme for energy storage units was derived. The focus was on the incorporation of stability constraints, based on Lyapunov theory and the concept of passivity. \\
It has been shown that the concept of passivity as well as the concept of Control Lyapunov Functions (CLF) can easily be merged with the idea of MPC. The stability properties of the different MPC setups were derived, implemented and simulated. Furthermore the corresponding control performance was analyzed. 

It was shown, that a standard MPC controller might evolve into unstable behavior. One of the reasons influencing this might be that energy storage charging level $x_{\textrm{SoC}}$ and the control input $u$ are penalized in the chosen setup. Augmenting this setup with stability constraints, such as in passivity based or CLF based MPC, stabilizes its behavior under otherwise equal conditions for all analyzed cases. To illustrate this in a study case, simulations were conducted for controlling a two area power system in a centralized and a decentralized setup. 
Using a longer prediction horizon might stabilize the conventional MPC controller, although this is usually not guaranteed. 

Although the computational demands for passivity based MPC might be comparably higher than for standard MPC at equal prediction horizons, a considerably longer prediction horizon seems to be needed for achieving (not guaranteeing) stability in a standard MPC setup. This could lead to less computational costs for the passivity based MPC approach, e.g. a prediction horizon of only $N\leq 3$ is needed in this setup with active passivity constraints to achieve acceptable control performance.\\
Regarding CLF based MPC, especially for the uncoordinated case, the time needed per optimization step is similar to the time needed within the standard MPC approach and results in guaranteed stability as in the passivity based approach.

Due to the fact, that asymptotic closed loop stability of the CLF based and passivity based MPC is guaranteed independently of the choice of $\mathcal{Q}$ and $\mathcal{R}$ in the quadratic objective functional, a separation between performance and stability considerations is achieved to some degree. Specifically, the tuning of the corresponding MPC cost weights might be carried out with less effort to achieve acceptable control performance and stability.


\subsection{Future Research}
\label{sec:future_research}

It might be promising to analyze how stability constraints (especially in a decentralized setup) would perform in a multi-node power network. It is shown in \cite{varaiya1983}, that for example in a three-node power network, the system may not be completely stable, even when the damping is made arbitrarily large -- unlike in a two-node power network.\\
Furthermore, investigations on robustness properties of the proposed control schemes as well as including both active and reactive power transmission into the corresponding models might be of interest.\\

\appendices





\ifCLASSOPTIONcaptionsoff
  \newpage
\fi



%




	\bibliographystyle{ieeetr}
	\bibliography{references}

\begin{thebibliography}{10}

\bibitem{res_infeed_germany}
``{Deutschland: Wind und Solar produzieren erstmals �ber 60 $\%$ des
  Stroms}.''
  \url{www.ee-news.ch/de/erneuerbare/article/26739/deutschland-wind-und-solar-produzieren-erstmals-ueber-6-des-stroms};
  Internationales Wirtschaftsforum Regenerative Energien (IWR), June 2013.

\bibitem{psd_andersson}
G.~Andersson, {\em Dynamics and Control of Electric Power Systems}.
\newblock EEH-Power Systems Laboratory, ETH Zurich, Lecture Notes, 2012.

\bibitem{paper_rinke}
T.~Rinke, A.~Ulbig, S.~Chatzivasileiadis, and G.~Andersson, ``Predictive
  control for real-time frequency control and inertia maximizing in power
  systems,'' {\em IEEE Transactions on Sustainable Energy}, 2013.

\bibitem{kundur1176power}
P.~Kundur, N.~Balu, and M.~Lauby, ``Power system stability and control, {EPRI}
  power system engineering series, 1994.''

\bibitem{psa_andersson}
G.~Andersson, {\em Power System Analysis}.
\newblock EEH-Power Systems Laboratory, ETH Zurich, Lecture Notes, 2012.

\bibitem{general_fc_ulbig}
A.~Ulbig, M.~Galus, S.~Chatzivasileiadis, and G.~Andersson, ``General frequency
  control with aggregated control reserve capacity from time-varying sources:
  The case of phevs,'' in {\em Bulk Power System Dynamics and Control
  (iREP)-VIII (iREP), 2010 iREP Symposium}, pp.~1--14, IEEE, 2010.

\bibitem{powernodes_ulbig}
K.~Heussen, S.~Koch, A.~Ulbig, and G.~Andersson, ``Unified system-level
  modeling of intermittent renewable energy sources and energy storage for
  power system operation,'' {\em Systems Journal, IEEE}, vol.~6, no.~1,
  pp.~140--151, 2012.

\bibitem{weissbach2008improvement}
T.~Wei{\ss}bach and E.~Welfonder, ``Improvement of the performance of scheduled
  stepwise power programme changes within the european power system,'' in {\em
  Proceedings of the 17th World Congress, The International Federation of
  Automatic Control (IFAC), Seoul, Korea}, pp.~11972--11977, 2008.

\bibitem{biasfactor}
M.~Scherer, E.~Iggland, A.~Ritter, and G.~Andersson, ``Improved frequency bias
  factor sizing for non-interactive control,'' {\em CIGRE 2012}, 2012.

\bibitem{ifac_welfonder}
M.~Kurth and E.~Welfonder, ``Importance of the selfregulating effect within
  power systems,'' {\em IFAC Symposium on Power Plants and Power Systems
  Control, Kananaskis, Canada}, pp.~345--352, 2006.

\bibitem{carpentier1985or}
J.~Carpentier, ``'to be or not to be modern' that is the question for automatic
  generation control (point of view of a utility engineer),'' {\em
  International Journal of Electrical Power \& Energy Systems}, vol.~7, no.~2,
  pp.~81--91, 1985.

\bibitem{survey_mpc}
D.~Mayne, J.~Rawlings, C.~Rao, and P.~Scokaert, ``Constrained model predictive
  control: Stability and optimality,'' {\em Automatica}, vol.~36, no.~6,
  pp.~789 -- 814, 2000.

\bibitem{sontag1989universal}
E.~D. Sontag, ``A universal construction of artstein's theorem on nonlinear
  stabilization,'' {\em Systems \& control letters}, vol.~13, no.~2,
  pp.~117--123, 1989.

\bibitem{clf_jadbabaie}
A.~Jadbabaie, J.~Yu, and J.~Hauser, ``Stabilizing receding horizon control of
  nonlinear systems: a control lyapunov function approach,'' in {\em American
  Control Conference, 1999. Proceedings of the 1999}, vol.~3, pp.~1535--1539,
  IEEE, 1999.

\bibitem{primbsdoyle}
J.~A. Primbs, V.~Nevisti{\'c}, and J.~C. Doyle, ``Nonlinear optimal control: A
  control lyapunov function and receding horizon perspective,'' {\em Asian
  Journal of Control}, vol.~1, no.~1, pp.~14--24, 1999.

\bibitem{passivity_raff}
T.~Raff, C.~Ebenbauer, and F.~Allg\"ower, ``Passivity-based nonlinear model
  predictive control,'' {\em Assessment and Future Directions of Nonlinear
  Model Predictive Control}, pp.~67--80, 2007.

\bibitem{sepulchre1997constructive}
R.~Sepulchre, M.~Jankovi{\'c}, and P.~Kokotovi{\'c}, {\em Constructive
  nonlinear control}.
\newblock Communications and control engineering, Springer, 1997.

\bibitem{khalil}
H.~K. Khalil and J.~Grizzle, {\em Nonlinear systems}, vol.~3.
\newblock Prentice hall Upper Saddle River, 2002.

\bibitem{freeman2008robust}
R.~A. Freeman and P.~V. Kokotovic, {\em Robust nonlinear control design:
  state-space and Lyapunov techniques}.
\newblock Birkhauser, 2008.

\bibitem{varaiya1983}
A.~Arapostathis and P.~Varaiya, ``Behaviour of three-node power networks,''
  {\em International Journal of Electrical Power \& Energy Systems}, vol.~5,
  no.~1, pp.~22--30, 1983.

\end{thebibliography}


%





\end{document}